\documentclass[journal,comsoc]{IEEEtran}

\usepackage{cite}
\usepackage{graphicx}
\usepackage{CJK}
\usepackage{amsmath}
\usepackage{amssymb}
\usepackage{textcomp}
\usepackage{pifont}
\usepackage{dblfloatfix}
\usepackage{cancel}
\usepackage[linewidth=0.5pt,linecolor=blue]{mdframed}

\newtheorem{theorem}{Theorem}

\newtheorem{lemma}{Lemma}{}
\newtheorem{corollary}{Corollary}{}
\newtheorem{remark}{Remark}{}
{}

\begin{document}

\title{
Discrete-Time Modeling and Handover Analysis of Intelligent Reflecting Surface-Assisted Networks
}


\markboth{IEEE TRANSACTIONS ON COMMUNICATIONS}%
{Shell \MakeLowercase{\textit{et al.}}: }

\author{  Haoyan~Wei,~\IEEEmembership{Member,~IEEE} and Hongtao~Zhang,~\IEEEmembership{Senior~Member,~IEEE,}
\thanks{
This work was supported in part by the National Natural Science Foundation of China under Grant 62271082, in part by Beijing Natural Science Foundation under Grants L242079, 4242007, and L234031, and in part by BUPT Excellent Ph.D. Student Foundation under Grant CX2023214.

H. Wei and H. Zhang are with Beijing University of Posts and Telecommunications, Beijing 100876, China (e-mail: weihaoyan@bupt.edu.cn; htzhang@bupt.edu.cn).
}
}

\maketitle

\IEEEpubid{\begin{minipage}{\textwidth}\ \centering
\\[1.5cm]
		© 2025 IEEE. Personal use of this material is permitted. Permission from IEEE must be obtained for all other uses, including reprinting/republishing this material for advertising or promotional purposes, collecting new collected works for resale or redistribution to servers or lists, or reuse of any copyrighted component of this work in other works.
\end{minipage}}

\begin{abstract}
Owning to the reflection gain and double path loss featured by intelligent reflecting surface (IRS) channels, handover (HO) locations become irregular and the signal strength fluctuates sharply with variations in IRS connections during HO, the risk of HO failures (HOFs) is exacerbated and thus HO parameters require reconfiguration. However, existing HO models only assume monotonic negative exponential path loss and cannot obtain sound HO parameters. This paper proposes a discrete-time model to explicitly track the HO process with variations in IRS connections, where IRS connections and HO process are discretized as finite states by measurement intervals, and transitions between states are modeled as stochastic processes. Specifically, to capture signal fluctuations during HO, IRS connection state-dependent distributions of the user-IRS distance are modified by the correlation between measurement intervals. In addition, states of the HO process are formed with Time-to-Trigger and HO margin whose transition probabilities are integrated concerning all IRS connection states. Trigger location distributions and probabilities of HO, HOF, and ping-pong (PP) are obtained by tracing user HO states. Results show IRSs mitigate PPs by 46\% but exacerbate HOFs by 91\% under regular parameters. Optimal parameters are mined ensuring probabilities of HOF and PP are both less than 0.1\%.
\end{abstract}
\vspace{-0.1cm}

\IEEEpeerreviewmaketitle

\vspace{-0.4cm}
\section{Introduction}
\vspace{-0.15cm}

\IEEEPARstart {I}{ntelligent} reflecting surface (IRS) has emerged as a promising solution to cope with the ever-growing demands of capacity through the new concept of reconfiguring the wireless propagation environment \cite{b1}. However, passive reflection gains from IRS cause handover (HO) locations to shift irregularly \cite{b1.1} and double path loss (also termed multiplicative fading) of the IRS channel causes the signal strength to drop steeply during user movement \cite{b1.2}, which poses challenges to HO performance.

IRS is typically a planar surface consisting of massive passive reflecting elements that can dynamically adjust the phase of the reflected signal to enhance the received signal strength from the serving base station (BS).\footnote{IRSs can also be non-planar, e.g., the conformal IRS discussed in \cite{b1.3}.}
However, signals from neighboring BSs are not specifically adjusted because no IRS is scheduled \cite{b2}, which creates distinct IRS reflection gains and shifts HO trigger locations.
As IRSs are deployed in a distributed manner \cite{b3,UAVIRS}, HO trigger locations become irregular, causing the risks of unreasonable HO triggers. 
Besides, the signal reflected by the IRS suffers double path loss as the signal passes through two paths \cite{b4}, where the IRS reflection gain decreases rapidly as the user moves away from the IRS. During the HO process, the sharp signal degradation caused by the double path loss brings the risk of handover failure (HOFs) \cite{b5}.
Moreover, the user connects, disconnects, and reselects the IRS while on the move \cite{b1.1}. Since the signal strength depends highly on the IRS connection, the dynamic IRS connection causes signal fluctuations and affects the HO process.

Due to the irregular HO trigger locations, sharp signal degradations, and fluctuations during the HO process introduced by the IRS, the HO parameters commonly used in traditional networks are no longer applicable. An analytical HO model is desired for sound HO parameter settings of IRS-assisted networks.
As for the analytical model, while previous works have comprehensively analyzed coverage performance with IRS (e.g.,\cite{b4.1} and \cite{b4.2}), the HO analysis of IRS-assisted networks is still in its infancy.

\subsection{Related Works}

Stochastic geometry has been extensively used to analyze HO performance metrics.
HO analysis begins with modeling the HO trigger location to evaluate the number of HOs. The HO trigger locations were modeled as sets of points equidistant from BSs in single-tier terrestrial networks \cite{b6,b7} and aerial networks \cite{b8}. But a closer BS may not provide a stronger signal, e.g., in heterogeneous networks, thus the user equipment (UE)-BS Euclidean distance is not sufficient to determine HO triggers. To address this issue, equivalent analysis techniques \cite{b9,b10} and analytical geometric frameworks \cite{b11,b12} were proposed to model HO trigger locations in heterogeneous networks. However, the IRS channel gain varies with the geometric relations of the IRS, BS, and UE, instead of linearly adjusting the signal strength, methods in  \cite{b9,b10,b11,b12} cannot obtain the exact HO trigger locations in IRS-assisted networks.

Furthermore, the number of HOs alone cannot reflect the HO performance, since HOFs and ping-pongs (PPs) have been found to cause severe disruptions and extra overhead [18]. Therefore, analyses of the HO process have emerged to enable HOF and PP evaluation and to guide HO parameter settings, i.e., time-to-trigger (TTT) and HO margin \cite{331}. The user sojourn time was analyzed in \cite{b14} and \cite{b15} to evaluate PPs, but the HOF analysis was lacking. HO models were proposed to derive both HOF and PP probabilities for single-tier networks \cite{SingleFad, Single}, heterogeneous networks \cite{Hetnet,HetnetFad}, and aerial networks \cite{UAVUser,UAVHO,UAVFading}.
For IRS-assisted networks, the study on HO performance is still in its infancy. The minimization of HOs was formed as an optimization problem in \cite{RISSF} rather than as an analytical model. \cite{IRSHO} performed preliminary analysis to obtain HO probabilities considering IRSs. 

Although several HO models were proposed and preliminary studies of the HO in IRS-assisted networks were conducted as mentioned, many issues remain unexplored in developing an analytical HO model for IRS-assisted networks, which are summarized as follows:
\begin{itemize}
\item The IRS channel gain with user movement has not been tracked theoretically. Negative exponential path-loss is adopted in \cite{Single, Hetnet, UAVUser,UAVHO} to characterize the trend in signal strength. However, IRS reselections cause signal fluctuations. Signal fluctuations were modeled as independent for each measurement in \cite{SingleFad, HetnetFad, UAVFading}. However, the relative location of the user to the IRS between measurements is correlated. IRS channels were introduced into HO probability analysis in \cite{IRSHO}. Nevertheless, \cite{IRSHO} only evaluated whether an HO is triggered within a unit of time, thus lacking IRS channel gain modeling during the whole HO process.

\item  The effect of the IRS on the HO process has not been introduced. Locations of HOF and PP are modeled as circular boundaries in \cite{Single, Hetnet, UAVUser,UAVHO,HetnetFad, UAVFading}, enabling HO process analysis. However, distributed IRSs make locations of HO events irregular, thus circular boundary models are ineffective. The transitions between HO states are modeled explicitly in \cite{SingleFad}. Nevertheless, owing to the lack of considering IRS channels, the HO state transitions in \cite{SingleFad} are not applicable for the HO process in IRS-assisted networks.

\item Reference-worthy results for the HO of IRS-assisted networks have not been derived. Lacking analysis of signal strength fluctuations and HO state transitions with IRS, models in \cite{SingleFad, Single,Hetnet, UAVUser,UAVHO,HetnetFad,UAVFading} are inaccurate in indicating actual HO performance and guiding HO parameters for IRS-assisted networks. Simulation-based or analytical model-based HO probabilities with IRS are given in \cite{RISSF, IRSHO}. Nevertheless, HO probabilities alone cannot represent the mobility performance, creating an urgent need to analyze both HOF and PP, and it is difficult to obtain guidelines via the computationally intensive simulation in \cite{RISSF}.

\end{itemize}

\begin{figure}[!t]
\vspace{-0.2cm}
\centerline{\includegraphics[width=0.9\linewidth]{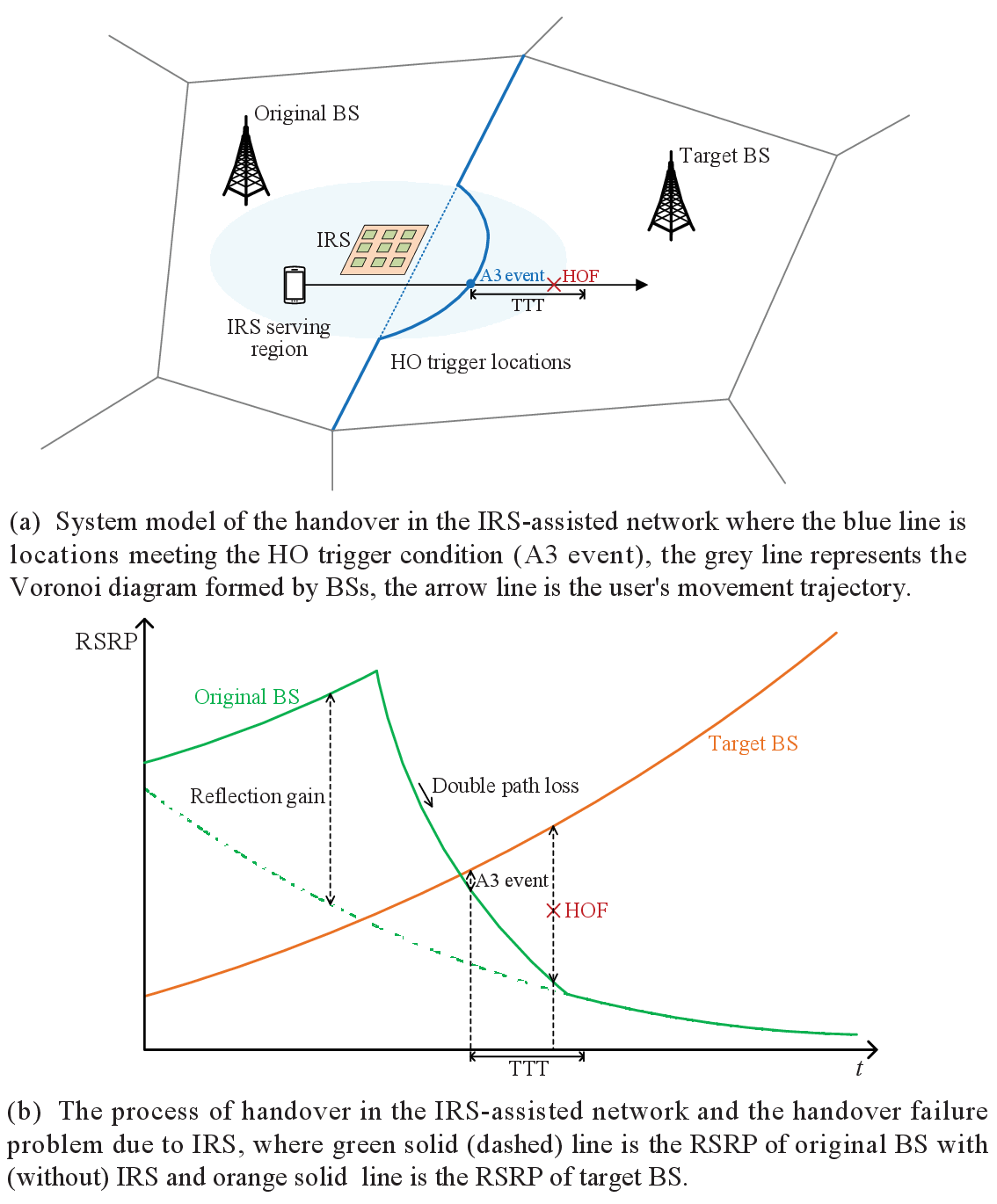}}
\vspace{-0.2cm}
\caption{Handover process in the IRS-assisted networks: (a) The IRS-assisted network structure; (b) The received signal strength changes with time.}
\vspace{-0.6cm}
\label{fig1}
\end{figure}

\vspace{-0.3cm}
\subsection{Contribution}

This paper proposes a discrete-time model to explicitly model signal strength fluctuations, enabling the HO analysis of IRS-assisted networks, where processes of IRS connection and HO are defined as finite states and tracked without  information loss. The main contributions are summarized as follows:

\begin{itemize}
\item The signal strength fluctuation during HO under the IRS cascaded channel is tracked theoretically. As the process of IRS connection is defined in the state space of the discrete-time model, the geometric relation between the HO user and its serving IRS is tracked. The probability density function (pdf) of IRS-user distance is revised via the correlation between measurement intervals to obtain the IRS reflection gain along the user trajectory.

\item The HO state transitions during the HO process are modified by IRS reflection gains. The processes for HO events (HO, HOF, and PP) occurring are described based on the actual timing relationships of the HO process and HO parameters. IRS reflection gains are introduced into transition probability matrices of HO states to analyze the exact HO trigger location, SIR changes within TTT duration, and user sojourn time.

\item Critical HO performance metrics for IRS-assisted networks are deduced.
HO state vectors are constituted by probabilities of HO states. HO state vectors along the user trajectory are obtained based on the discrete-time model of the IRS connection and HO process. Compact expressions of HOF and PP probability are deduced by extracting elements of the HO state vectors, and the trigger location distributions of HO events are obtained.

\item Design insights obtained from the main results include: (i) The IRS implementation provides a new tradeoff for HOF and PP, where IRSs mitigate the PP probability by 46\% but exacerbate the HOF probability by 91\% when the density of IRS is 500/km$^2$, and the element number is 100; (ii) Guidance for the setting of HO parameters (TTT and HO margin) is studied, optimal HO parameters are mined to make the probability of PP and HOF both less than 0.1\% under multiple network setups. 
\end{itemize}

The paper is structured as follows: Section II outlines the system model. Section III characterizes the discrete-time model for IRS and HO. Section IV presents and discusses numerical results for HO, HOF, and PP events. Key notations are summarized in Table I.

\section{System Model}

\subsection{Network Model}
\vspace{-0.1cm}
In this paper, we consider an IRS-assisted wireless network, as shown in Fig 1(a).\footnote{For brevity, only one IRS that affects the HO process is shown in Fig 1(a), and other IRSs are omitted. In the network considered, each BS controls multiple IRSs in its cell and the HO process may involve several IRSs.} 
Assume for simplicity that the BSs and UEs are each equipped with a single antenna, while each IRS has $N$ elements. 
BSs are distributed according to a 2D homogenous Poisson point process (HPPP) with density $\lambda_b$ and the same transmitting power $P_t$. IRSs are distributed according to a 2D HPPP with density $\lambda_r$ \cite{b4}.\footnote{The HPPP assumption is universal and widely accepted for modeling locations of BSs and IRSs in IRS-assisted networks \cite{b4, b4.1, b4.2}. Therefore special point processes are not adopted in this paper, in order to maintain the focus on the most fundamental aspects of HO analysis.} Each IRS is controlled by its nearest BS \cite{b3}, where the BS controls IRSs in its cell via wired/wireless control links. For a user, its connected BS schedules an IRS in its cell that provides the highest gain (i.e., the IRS nearest to the user) for dedicated  reflect beamforming.
A limited IRS serving distance is considered, as the IRS provides signal enhancement in a \emph{local} region \cite{b4, b26}, therefore, the user will not be served by the IRS if there is no IRS within a threshold distance.
 
According to 3GPP specifications \cite{331}, the UE performs channel measurements periodically, and the measurement interval is defined as $T_d$.
Owing to the filtering process at UE, the effect of the fast-fading of the channel is averaged out \cite{Single, Hetnet}. 
To capture the IRS reflection gain, the channel model proposed in [7] is adopted, where the serving IRS provides a reflective path in addition to the direct BS-user path to enhance the signal strength.
For the BS to which the user is connected before HO, its signal is enhanced by an IRS if it exists within the limited IRS serving distance from the user; thus, the received signal strength is given by \cite{b4}
\begin{equation}
\setlength\abovedisplayskip{3pt}
\setlength\belowdisplayskip{3pt}
{{\cal S}^c}\left( {x,d,\varphi } \right) = \left\{ {\begin{array}{ll}
{{P_t}{\Gamma _{bf}}\left( {x,d,\varphi } \right),}&{d \le D}\\
{{P_t}{\Gamma _{sc}}\left( {x,d,\varphi } \right),}&{d > D}
\end{array}} \right.,
\end{equation}
where $x$ and $d$ denote user-BS distance and user-IRS distance, $\varphi$ denotes the BS-user-IRS angle, $D$ is the limited IRS serving distance, ${\Gamma _{bf}}\left( {x,d,\varphi } \right)$ and ${\Gamma _{sc}}\left( {x,d,\varphi } \right)$ indicate path losses in the case of IRS dedicated reflection and IRS random scattering, which are given by
\begin{equation}
\setlength\abovedisplayskip{3pt}
\setlength\belowdisplayskip{3pt}
\begin{small}
\begin{aligned}
{\Gamma _{bf}}\left( {x,d,\varphi } \right) &= {g_b}\left( x \right) + {G_{bf}}{g_r}\left( d \right){g_b}\left( {x'} \right)\\
 &+ N\frac{\pi }{4}\sqrt {\pi {g_b}\left( x \right){g_r}\left( d \right){g_b}\left( {x'} \right)},\\[-1mm]
{\Gamma _{sc}}\left( {x,d,\varphi } \right) &= {g_b}\left( x \right) + N{g_b}\left( {x'} \right){g_r}\left( d \right),
\end{aligned}
\end{small}
\end{equation}
where $g_b{\left(x \right)}=\beta {x^{ - \alpha }}$, $g_r{\left(d\right)}=\beta {d^{ - \alpha }}$, ${G_{bf}} = \frac{{{\pi ^2}}}{{16}}{N^2} + \left( {1 - \frac{{{\pi ^2}}}{{16}}} \right)N$, $x' = \sqrt {{x^2} + {d^2} - 2xd\cos \varphi } $ is the BS-IRS distance, ${ \alpha }$ is the path-loss exponent, $\beta  = {\left( {{{4\pi {f_c}} \mathord{\left/
 {\vphantom {{4\pi {f_c}} c}} \right.
 \kern-\nulldelimiterspace} c}} \right)^{ - 2}}$, $f_c$ is the carrier frequency, $c$ is the speed of light.\footnote{Like the scenario of interest in most works (e.g.,\cite{b3, UAVIRS, b4, b4.1, b4.2}), we consider each user is served by at most one IRS, and scattering component of other IRSs is also omitted due to the fact that it is extremely small \cite{b4}.}

For the neighboring BS, because no IRS reflecting beamforming is provided for the measured reference signal \cite{b4, IRSHO}, the received signal strength is ${{\cal S}^n}\left( {x,d,\varphi } \right) = {P_t}{\Gamma _{sc}}\left( d \right)$.

\begin{remark}
Although the considered BS and UE are equipped with a single antenna, the proposed model can be extended to multi-antenna systems with minor modifications, e.g., by multiplying ${G_T}\left( {\phi _b^T} \right){G_R}\left( {\phi _b^R} \right)$, ${G_T}\left( {\phi _r^T} \right)$, ${G_R}\left( {\phi _r^R} \right)$ before ${g_b}\left( x \right)$, ${g_b}\left( {x'} \right)$, ${g_r}\left( d \right)$ respectively in (2), where ${\phi _b^T}$ (${\phi _r^T}$) is angle of departure  from the BS to the user (IRS), ${\phi _b^R}$ (${\phi _r^R}$) is angle of arrival from the BS (IRS) to the user, ${G_T}\left( {\phi} \right)$ and ${G_R}\left( {\phi} \right)$ are array gain functions of BS and UE set as need.
\end{remark}
\begin{remark}
The IRS serving distance $D$ implies that the IRS serves the user if the relative signal strength gain reaches a threshold, where the threshold is ${{\gamma _{{\rm{IRS}}}}} = \frac{{{P_t}{\Gamma _{bf}}\left( {x,D,\varphi } \right) }}{{{P_t}{\Gamma _{sc}}\left( {x,D,\varphi } \right)}}-1$.
The relation between $D$ and ${\gamma _{{\rm{IRS}}}}$ is \cite{b26}
\begin{equation}
\setlength\abovedisplayskip{2pt}
\setlength\belowdisplayskip{2pt}
\begin{small}
\begin{aligned}
 D \approx \sqrt[\alpha ]{{\frac{{{G_{bf}}\beta }}{{{\gamma _{{\rm{IRS}}}}}}}},
\end{aligned}
\end{small}
\end{equation}
which holds approximately due to $\!{g_b}\!(x') \!\approx\! {g_b}\!(x)$ and $N{g_r}\!\!\left( D \right) \!\!\ll\! 1$. $D$ can be determined by pragmatically setting ${\gamma _{{\rm{IRS}}}}$. Other IRS access strategies can be applied to the proposed model by modifying $D$ as functions, random variables, etc.
\end{remark}


\begin{table}
\renewcommand\arraystretch{1.05}
\vspace{-0.1cm}
	\caption{ {Table of Key Notations}}               
	\begin{center}
\vspace{-0.5cm}
		\begin{tabular}{|c ||c|} 
\hline \bf{Notation} &\bf{Description}\\    
\hline 	$\lambda_b$, $\lambda_r$ & Densities of BS and IRS \\
\hline   $N$                     & Number of IRS elements    \\
\hline   $D$                     & IRS serving diatance    \\
\hline   ${\Gamma _{bf}}\left( {x,d,\varphi } \right)$, 
                                 & Path losses under IRS dedicated reflection \\[-0.5mm] 
${\Gamma _{sc}}\left( {x,d,\varphi } \right)$ & and IRS random scattering    \\
\hline   $x_o$, $x_t$, $d$       & Distances of the user from the original BS, \\[-1mm] &the target BS, and the IRS    \\
\hline   ${\varphi_o}$, ${\varphi_t}$, ${\varphi '}$  & BS-user-IRS angles of the original and target \\[-1mm] & BS, and angle from the trajectory to the IRS \\
\hline   $T_t$, ${\gamma _{{\rm{HO}}}}$ & TTT and HO margin   \\
\hline   $Q_{out}$, $T_p$             & Thresholds for HOF and PP    \\
\hline   $\mathcal{I}_1^k$, $\mathcal{I}_2^k$, $\mathcal{I}_3^k$, $\mathcal{I}_4^k$            & IRS connection states    \\
\hline   ${\mathcal{H}}_0$, $\cdots$, ${\mathcal{H}}_j$, ${\mathcal{H}}_c$            & States of the HO process   \\
\hline   ${\mathcal{F}}_0$, $\cdots$, ${\mathcal{F}}_j$, ${\mathcal{F}}_t$, ${\mathcal{F}}_c$            & States of the HOF   \\
\hline   \!\!${\mathcal{P\!P}}\!_0$, \!\!$\!\cdots$\!, \!\!${\mathcal{P\!P}}\!\!_u$, \!${\mathcal{P\!P}}\!\!_t$, \!${\mathcal{P\!P}}\!\!_c$ \!\!           & States of the PP event   \\
\hline ${{\rm{\bf{T}}}^{{\mathcal I}^k}}\!\!\!\left( i \right)$, ${{\rm{\bf{T}}}^{\mathcal{H}}}\!\left( i \right)$, & State transition matrices of IRS connection, \\[-0.7mm] ${{\rm{\bf{T}}}^{\mathcal{F}}}\!\left( i \right)$, ${{\rm{\bf{T}}}^{\mathcal{PP}}}\!\left( i \right)$ & HO,  HOF, and PP    \\
\hline ${{\rm{\bf{S}}}^{{\mathcal I}^k}}\!\!\!\left( i \right)$, ${{\rm{\bf{S}}}^{\mathcal{H}}}\!\left( i \right)$, & State vectors of IRS connection, HO, HOF,\\[-0.7mm] ${{\rm{\bf{S}}}^{\mathcal{F}}}\!\left( i \right)$, ${{\rm{\bf{S}}}^{\mathcal{PP}}}\!\left( i \right)$ & and PP    \\
\hline   ${{\mathbb{P}}_{ht}}\!\left( {{x_i}} \right)$ , ${{\mathbb{P}}_{ho}}\!\left( {{x_i}} \right)$                   & Probabilities of HO trigger and HO execution \\[-1mm]
& at $i$-th measurement \\
\hline   ${{\mathbb{P}}_{hof}}$ , ${{\mathbb{P}}_{pp}}$                   & Probabilities of HOF and PP\\
\hline ${\mathbb{E}}\left[ {{x^{ht}}} \right] $, ${\mathbb{E}}\left[ {{x^{ho}}} \right] $, & \!\!\!Average distances for HO trigger, HO execution,\!\!\! \\[0.1mm] ${\mathbb{E}}\left[ {{x^{hof}}} \right] $, ${\mathbb{E}}\left[ {{x^{pp}}} \right] $ & HOF, and PP    
  \\ \hline
		\end{tabular}
	\end{center}
\vspace{-0.6cm}
\end{table}

\subsection{IRS Connection Events}

In IRS-assisted networks, the BS dynamically schedules the best IRS for the user to perform the service (if any IRSs of that cell exists within a distance $D$ from the user) based on the measurement \cite{b4, IRSHO, b26}. Therefore, the following IRS connection events are considered.

\emph{Initial connection to IRS}: The user not served by the IRS steps into the serving region of an IRS of the serving cell. Alternatively, the user handovers to a new cell and connects to an IRS of the new cell.

\emph{Disconnection of IRS}: The user leaves its IRS serving region and there is no IRS within $D$ of the serving cell. Alternatively, when the user handovers to a new cell, it disconnects from the IRS of the original cell.

\emph{Reselection}: When there is an IRS of the serving cell that is closer to the user than the serving IRS, the BS schedules that IRS to serve the user.

\vspace{-0.4cm}
\subsection{Handover Process and Events}


As shown in Fig. 1(b), the HO process involves the following steps:

\emph{HO trigger}: The A3 event is adopted to decide the HO trigger as in  \cite{SingleFad, Single, Hetnet,UAVUser,UAVHO, HetnetFad,UAVFading}. The A3 event is triggered when the received signal strength of the target BS becomes better than the serving BS, based on a certain offset, i.e., the HO margin.

\emph{TTT Timing}: When the condition of the A3 event is satisfied, the TTT timer starts, which is set to avoid unnecessary HOs.

\emph{HO execution}: Only when the condition of the A3 event is satisfied in the entire TTT duration, the UE feeds back the measurement report to the serving BS and waits for the HO command to perform subsequent steps. 

Therefore, the condition for HO execution is expressed as
\begin {equation}
\setlength\abovedisplayskip{3pt}
\setlength\belowdisplayskip{3pt}
\frac{{{{\mathcal S}^n}\!\left( {{x_t},d,{\varphi _t}} \right)}}{{{{\mathcal S}^c}\!\left( {{x_o},d,{\varphi _o}} \right)}} > {\gamma _{{\rm{HO}}}},\forall t \in \left[ {{t_0},{t_0} + {T_t}} \right],
\end {equation}
where $x_o$ and $x_t$ are the distances of the user from the original BS and the target BS, ${{\varphi _o}}$ and ${{\varphi _t}}$ are original BS-user-IRS angle and target BS-user-IRS angle, ${\gamma _{{\rm{HO}}}}$ is the HO margin, $t_0$ is the moment when the HO trigger condition is satisfied.

To evaluate the HO performance, the following two events are of most interest:

\emph{HOF}: A drop in wideband SIR to a threshold results in the inability to receive HO commands, i.e., HOF \cite{331}. As \cite{SingleFad, Single, Hetnet,UAVUser,UAVHO, HetnetFad,UAVFading}, an interference-limited system is considered, along with the interference from the target BS, which is the absolute main part. Thus, the condition of HOF is given by
\begin {equation}
\setlength\abovedisplayskip{3pt}
\setlength\belowdisplayskip{3pt}
\frac{{{{\mathcal S}^c}\!\left( {{x_o},d,{\varphi _o}} \right)}}{{{{\mathcal S}^n}\!\left( {{x_t},d,{\varphi _t}} \right)}} < {Q_{out}},\exists t \in \left[ {{t_0},{t_0} + {T_t}} \right].
\end {equation}
where $Q_{out}$ denotes the threshold for HOF.

\emph{PP}: The occurrence of PP depends on the user sojourn time in the new cell. If the user handovers to a new BS and handovers back to the original BS within a certain time threshold, the HO is considered unnecessary, which is defined as a PP. After HO, the user disconnects from the original BS and is connected to the target BS. Thus, a PP occurs when conditions (3) and (5) are satisfied, and (5) is given by
\begin {equation}
\setlength\abovedisplayskip{3pt}
\setlength\belowdisplayskip{3pt}
\frac{{{{\mathcal S}^n}\!\left( {{x_o},d,{\varphi _o}} \right)}}{{{{\mathcal S}^c}\!\left( {{x_t},d,{\varphi _t}} \right)}} > {\gamma _{{\rm{HO}}}},\exists t \in \left[ {{t_0} + {T_t},{t_0} + {T_p}} \right],
\end {equation}
where $T_p$ denotes the minimum threshold for the sojourn time. 

\vspace{-0.4cm}
\subsection{User Mobility}
\vspace{-0.1cm}

Without loss of generality, a typical user is considered to move at a speed $v$ and direction during the HO process, which is the same as the assumptions in \cite{SingleFad, Single, Hetnet,UAVUser,UAVHO, HetnetFad,UAVFading}. The BSs involved in the HO process are considered, which are the original BS and the target BS. The location relations are shown in Fig. 2. The distance from the original BS (target BS) to the user trajectory is denoted as $r_o$ ($r_t$); in particular, $r_t$ takes a negative value when BSs are on both sides of the trajectory. We focus on the part between the foot points of two BSs on the trajectory, whose length is denoted as $L$. The probability density function (pdf) of $L$ is given by $\rm{[33,\ Eq. (14)]}$
\begin {equation}
\setlength\abovedisplayskip{1pt}
\setlength\belowdisplayskip{1pt}
\begin{footnotesize}
\begin{aligned}
{f_L}\left( l \right) =  \int\limits_0^\pi  {\int\limits_0^\pi  {\frac{{{\pi ^2}\lambda _b^{\frac{3}{2}}{l^2}{\rho _\upsilon }{\rho _\tau }\left( {2\pi {\lambda _b}{l^2}b_0^2\rho _\upsilon ^2 - {c_0}} \right)}}{{\sin \left( {\upsilon  + \tau } \right)}}} }& \\[-3mm]
 \times \exp \left( { - \pi {\lambda _b}{l^2}{V_2}\left( {\upsilon ,\tau } \right)} \right){\rm{d}}\upsilon {\rm{d}}\tau , &
\end{aligned}
\end{footnotesize}
\end {equation}
where ${\rho _\upsilon } = \frac{{\sin \upsilon }}{{\sin \left( {\upsilon  + \tau } \right)}}$, ${\rho _\tau } = \frac{{\sin \tau }}{{\sin \left( {\upsilon  + \tau } \right)}}$, ${V_2}\left( {\upsilon ,\tau } \right) = \left( {1 + \rho _\tau ^2-2{\rho _\tau }\cos \upsilon } \right)\left( {1 - \frac{\tau }{\pi } + \frac{{\sin 2\tau }}{{2\pi }}} \right) + \rho _\tau ^2\left( {1 - \frac{\upsilon }{\pi } + \frac{{\sin 2\upsilon }}{{2\pi }}} \right)$, ${b_0} = \frac{{\left( {\pi  - \tau } \right)\cos \tau  + \sin \tau }}{\pi }$, and ${c_0} = \frac{{\left( {\pi  - \tau } \right) + \sin \tau \cos \tau }}{\pi }$. As in [22], $L$ is simplified to its expected value $\mathbb{E} \left[ L \right]$ to simplify the analysis and concentrate on the primary contribution. 
Define the foot point of the original BS as the initial location, the user trajectory is discretized based on the measurement interval $T_d$; thus, the user moves ${\Delta x}\!\!=\!{T_d}\cdot v$ in each measurement interval and is ${x_i} \!=\! i\cdot{\Delta x}$ from the initial location at $i$-th measurement, where $i \in \left\{ {0,1, \cdots ,I} \right\},I = \left\lceil {\frac{L}{{\Delta x}}} \right\rceil $ and $\left\lceil \cdot \right\rceil $ represents the ceiling function. The distances of the user from the original BS, the target BS, and the IRS are ${x_{o}} \!=\! \sqrt {x_i^2 + r_o^2} $, ${x_{t}} \!=\! \sqrt {{{\left( {L - {x_i}} \right)}^2} + r_t^2} $ and $d$. ${\varphi '}$ is the angle from the trajectory to the IRS, then original BS-user-IRS angle and target BS-user-IRS angle are ${\varphi _o} \!\!=\! \pi  \!-\! \arctan \frac{{{r_o}}}{{{x_i}}} - \varphi '$ and ${\varphi _t} \!=\! \varphi ' \!-\! \arctan \frac{{{r_t}}}{{{x_i}}}$. In addition, ${x_{{\rm{mid}}}} = {{\left( { {r_o^2 - r_t^2} + {L^2}} \right)}\mathord{\left/
\right.\kern-\nulldelimiterspace} 2L}$ is the distance between the initial location and the location on the trajectory that are equidistant from the BSs, $\theta $ is the angle between the trajectory and the line equidistant from the BSs, which are used for subsequent derivations.

\begin{figure}[!t]
\vspace{-0.3cm}
\centerline{\includegraphics[width=0.9\linewidth]{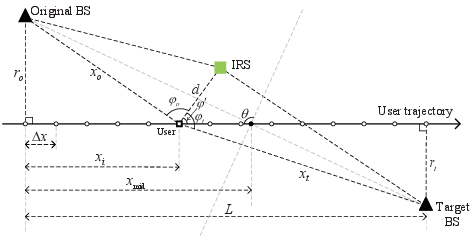}}
\vspace{-0.3cm}
\caption{User mobility mode and location relations between users and network elements.}
\vspace{-0.6cm}
\label{fig1}
\end{figure}

 \vspace{-0.1cm}

\begin{figure*}[!t]
\vspace{-0.5cm}
\centerline{\includegraphics[width=420pt]{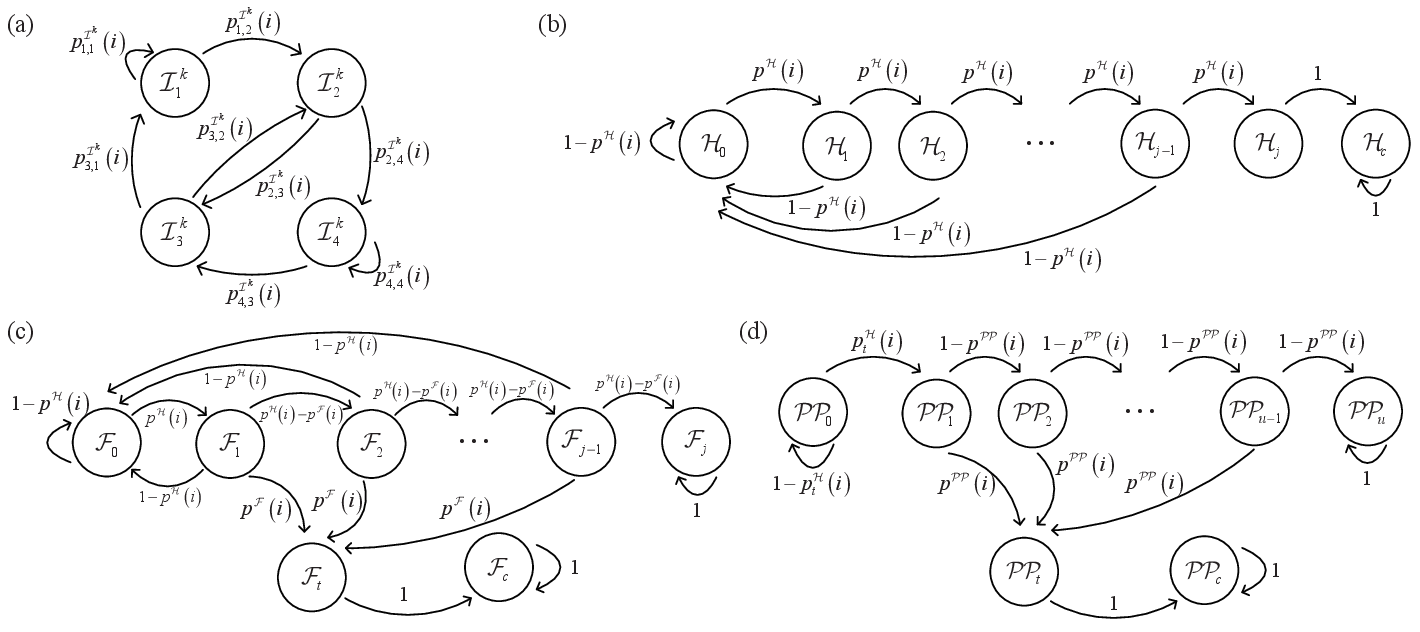}}
\vspace{-0.3cm}
\caption{Discrete-time models for (a) IRS connections, (b) handover, (c) handover failure, and (d) ping-pong. The symbols of each state are inside the circles and the state transition probabilities are beside the lines.}
\label{fig1}
\vspace{-0.6cm}
\end{figure*}

\section{Discrete-Time Models for IRS Connections and Handover}

 \vspace{-0.1cm}

\subsection{IRS Connections}
 
To track the signal strength fluctuations along the user trajectory theoretically, a discrete-time model of the IRS connection is established, as shown in Fig. 3(a). 
The correlation between measurements is introduced as the IRS connection at the previous measurement moment is always regarded.
Four IRS connection states are defined:  ${\mathcal I}^k_{1}$, ${\mathcal I}^k_{2}$, ${\mathcal I}^k_{3}$, and ${\mathcal I}^k_{4}$, $k \in \left\{ {o,t} \right\}$ represents the IRS connection state of IRSs controlled by the original BS or the target BS, and the specific definitions of the four IRS connection states are as follows:
\begin{enumerate}
\item \emph{No connection to IRS}: the user is not served by any IRSs in both moments, which is indicated by  ${\mathcal I}^k_{1}$.

\item \emph{Initial connection to IRS}: same as that described in Section II.B, which is indicated by  ${\mathcal I}^k_{2}$.

\item \emph{Disconnection of IRS}: same as that described in Section II.B, which is indicated by  ${\mathcal I}^k_{3}$.

\item \emph{Keeping the connection to IRS}: the user is consistently served by an IRS at two moments or the user is served by a closer IRS (i.e., reselection described in Section II.B), which is indicated by ${\mathcal I}^k_{4}$.

\end{enumerate}

Based on the definitions of the IRS connection states, the corresponding state transition matrix and state vector are constructed as follows.

\begin{lemma}
The state transition matrix of the IRS connection is given by
\begin{equation}
\setlength\abovedisplayskip{1pt}
\setlength\belowdisplayskip{3pt}
\begin{small}
{{\rm{\bf{T}}}^{{\mathcal I}^k}}\!\!\!\left( i \right) \!=\! \left[ {\begin{array}{*{20}{c}}
{p_{1,1}^{{{\mathcal I}^k}}\left( i \right)}&{p_{1,2}^{{{\mathcal I}^k}}\left( i \right)}&0&0\\
0&0&{p_{2,3}^{{{\mathcal I}^k}}\left( i \right)}&{p_{2,4}^{{{\mathcal I}^k}}\left( i \right)}\\
{p_{3,1}^{{{\mathcal I}^k}}\left( i \right)}&{p_{3,2}^{{{\mathcal I}^k}}\left( i \right)}&0&0\\
0&0&{p_{4,3}^{{{\mathcal I}^k}}\left( i \right)}&{p_{4,4}^{{{\mathcal I}^k}}\left( i \right)}
\end{array}} \right],
\end{small}
\end{equation}
where ${p_{m,n}^{{{\mathcal I}^k}}\left( i \right)}, k \in \left\{ {o, t} \right\}, m, n \in \left\{ {1,2,3,4} \right\}$ is state transition probabilities of IRS connection sates. ${p_{m,n}^{{{\mathcal I}^o}}\left( i \right)}$ and ${p_{m,n}^{{{\mathcal I}^t}}\left( i \right)}$ are given by
\begin{equation}
\setlength\abovedisplayskip{1pt}
\setlength\belowdisplayskip{3pt}
\begin{small}
\begin{aligned}
&p_{1,1}^{{{\cal I}^o}}\!\!\left( i \right) \!=\! p_{3,1}^{{{\cal I}^o}}\!\!\left( i \right) \!=\! \left\{ \begin{array}{ll}
{e^{ - {\lambda _r}\left( {S_i^o - S_i^{o, ov }} \right)}},&0 \le {x_i} \le {x_{{\rm{mid}}}} + \frac{D}{{\sin \theta }}\\
1,&{x_{{\rm{mid}}}} + \frac{D}{{\sin \theta }} < {x_i}
\end{array} \right.,\\
&p_{2,3}^{{{\cal I}^o}}\!\!\left( i \right) \!=\! p_{4,3}^{{{\cal I}^o}}\!\!\left( i \right) \!=\! \left\{ \begin{array}{ll}
\!\!\! \left( {1 \!-\! \frac{{{e^{ - {\lambda _r}S_i^{o, \cap }}}}}{{{e^{ - {\lambda _r}S_{i - 1}^o}}}}} \right)\!{e^{ \!\!- {\lambda _r}\left( {S_i^o - S_i^{o, \cap }} \right)}},\!\!\!\!\!\!& 0\! \le\! {x_i} \!\le\! {x_{{\rm{mid}}}} \!+\! \frac{D}{{\sin \theta }}\\
\!\!\! 1,&{x_{{\rm{mid}}}} + \frac{D}{{\sin \theta }} < {x_i}
\end{array} \right.\!\!,\\
& p_{1,2}^{{{\cal I}^o}}\!\!\left( i \right) \!=\! p_{3,2}^{{{\cal I}^o}}\!\!\left( i \right) \!=\! 1 - p_{1,1}^{{{\cal I}^o}}\left( i \right),\ \ p_{2,4}^{{{\cal I}^o}}\left( i \right) = p_{4,4}^{{{\cal I}^o}}\left( i \right) = 1 - p_{2,3}^{{{\cal I}^o}}\left( i \right),
\end{aligned}
\end{small}
\end{equation}
\begin{equation}
\setlength\abovedisplayskip{1pt}
\setlength\belowdisplayskip{3pt}
\begin{small}
\begin{aligned}
&p_{1,1}^{{{\cal I}^t}}\!\!\left( i \right) \!= p_{3,1}^{{{\cal I}^t}}\left( i \right) = \left\{ \begin{array}{ll}
1,&0 \le {x_i} \le {x_{{\rm{mid}}}} - \frac{D}{{\sin \theta }}\\
{e^{ - {\lambda _r}\left( {S_i^t - S_i^{t, \cap }} \right)}},&{x_{{\rm{mid}}}} - \frac{D}{{\sin \theta }} < {x_i}
\end{array} \right.,\\
&p_{2,3}^{{{\cal I}^t}}\!\!\left( i \right) \!=\! p_{4,3}^{{{\cal I}^t}}\!\!\left( i \right) \!= \! \left\{ \begin{array}{ll}
\!\!\!1,&\!\!\!\!\!\!\!\!\!\!\!\!\! 0 \le {x_i} \le {x_{{\rm{mid}}}} - \frac{D}{{\sin \theta }}\\
\!\!\!\left( {1 \!-\! \frac{{{e^{ - {\lambda _r}S_i^{t, \cap }}}}}{{{e^{ - {\lambda _r}S_{i - 1}^t}}}}} \right)\!{e^{ \!\! - {\lambda _r}\left( {S_i^t - S_i^{t, \cap }} \right)}},&\!\!\!\!\!{x_{{\rm{mid}}}} - \frac{D}{{\sin \theta }} < {x_i}
\end{array} \right.\!\!\!\!\!,\\
& p_{1,2}^{{{\cal I}^t}}\!\!\left( i \right) \!=\! p_{3,2}^{{{\cal I}^t}}\!\!\left( i \right) \!=\! 1 - p_{1,1}^{{{\cal I}^t}}\left( i \right),\ \ p_{2,4}^{{{\cal I}^t}}\left( i \right) = p_{4,4}^{{{\cal I}^t}}\left( i \right) = 1 - p_{2,3}^{{{\cal I}^t}}\left( i \right),
\end{aligned}
\end{small}
\end{equation}
 where $S_i^o$ ($S_i^t$) is the area of the region where the IRS of original (target) BS may exist at the $i$-th measurement, $S_i^{o, \cap}$ ($S_i^{t, \cap}$) is the area of overlap of the regions where the IRS of original (target) BS may exist at $(i-1)$-th and $i$-th measurements, and areas are given by
\begin{equation}
\setlength\abovedisplayskip{1pt}
\setlength\belowdisplayskip{3pt}
\begin{footnotesize}
\begin{aligned}
&S_i^o = \left\{ \begin{array}{ll}
\pi {D^2},&0 \le {x_i} \le {x_{{\rm{mid}}}} - \frac{D}{{\sin \theta }}\\
\pi {D^2} - {D^2}\left( {{\varepsilon _i} - \sin {\varepsilon _i}} \right)/2,&{x_{{\rm{mid}}}} - \frac{D}{{\sin \theta }} < {x_i}
\end{array} \right.,\\
&S_i^t = \left\{ \begin{array}{ll}
{D^2}\left( {{\varepsilon _i} - \sin {\varepsilon _i}} \right),&{x_{{\rm{mid}}}} - \frac{D}{{\sin \theta }} \le {x_i} \le {x_{{\rm{mid}}}} + \frac{D}{{\sin \theta }}\\
\pi {D^2},&{x_{{\rm{mid}}}} + \frac{D}{{\sin \theta }} < {x_i}
\end{array} \right.,\\
&S_i^{o, \cap } \!=\! \left\{ \begin{array}{l}
\!\!\! 2{D^2}\arccos \left( {\frac{{\Delta x}}{{2D}}} \right) \!-\! \frac{{\Delta x}}{2}\sqrt {4{D^2} - \Delta {x^2}} ,\ \ 0 \!\le\! {x_i} \!\le\! {x_{{\rm{mid}}}} \!-\! \frac{D}{{\sin \theta }} \!+\! \Delta x\\
\!\!\! \int\limits_0^D \! \int\limits_0^{2\pi }\!\! {\Delta S_i^{s, \cap }{\left( {\phi ,\rho } \right)} \rho {\rm{d}}\phi {\rm{d}}\rho } ,\ \ \ \ \ {x_{{\rm{mid}}}} \!-\! \frac{D}{{\sin \theta }} \!+\! \Delta x \!<\! {x_i} \!\le\! {x_{{\rm{mid}}}} \!+\! \frac{D}{{\sin \theta }}
\end{array} \right.,
\end{aligned}
\end{footnotesize}\notag
\end{equation}
\begin{equation}
\setlength\abovedisplayskip{2pt}
\setlength\belowdisplayskip{3pt}
\begin{footnotesize}
\begin{aligned}
&\Delta S_i^{o, \cap }\! {\left( {\phi ,\rho } \right)} \!=\! \left\{ \begin{array}{l}
\!\!\! 1,\ \ \left[ {\rho\! \sin \phi  \!+\! \rho \tan \!\theta \cos \!\phi  \!-\! \tan \!\theta \left( {{x_{{\rm{mid}}}} - {x_i}} \right)} \right]\left( {\theta  - \frac{\pi }{2}} \right) \!> 0\\ \ \ \ \ \ \ \ \ \ \ \ \ \ \ \ \ \ \ \ \ \ \ \ \ \  \wedge \  0 \!<\! \sqrt {\Delta {x^2} + {\rho ^2} + 2\rho \Delta x\cos \phi }  \!<\! D\\
\!\!\!0,\ \ \ {\rm{others}}
\end{array} \right.\!\!\!\!,\\
&S_i^{t, \cap } \!=\! \left\{ \begin{array}{lr}
\!\!0,&{x_{{\rm{mid}}}} - \frac{D}{{\sin \theta }} \le {x_i} \le {x_{{\rm{mid}}}} - \frac{D}{{\sin \theta }} + \Delta x\\
\!\!\int\limits_0^D  \! \int\limits_0^{2\pi } \!\!{\Delta S_i^{t, \cap }{\left( {\phi ,\rho } \right)} \rho {\rm{d}}\phi {\rm{d}}\rho } ,&{x_{{\rm{mid}}}} - \frac{D}{{\sin \theta }} + \Delta x < {x_i} \le {x_{{\rm{mid}}}} + \frac{D}{{\sin \theta }}\\
\!\!2{D^2}\arccos \left( {\frac{{\Delta x}}{{2D}}} \right) - \frac{{\Delta x}}{2}\sqrt {4{D^2} - \Delta {x^2}} ,\!\!\!\!\!\!\!\!\!\!\!\!\!\!\!\!\!\!\!\!\!\!\!\!\!\!\!\!\!\!\!\!\!\!\!\!\!\!\!\!\!\!\!\!\!\!\!&{x_{{\rm{mid}}}} + \frac{D}{{\sin \theta }} < {x_i}
\end{array} \right.\!\!\!,\\
&\Delta S_i^{t, \cap } \! {\left( {\phi ,\rho } \right)} \!=\! \left\{ \begin{array}{l}
\!\!\! 1,\ \ \left[ {\rho\! \sin \phi  \!+\! \rho \tan \!\theta \cos \!\phi  \!-\! \tan \!\theta \left( {{x_{{\rm{mid}}}} - {x_i}} \right)} \right]\left( {\theta  - \frac{\pi }{2}} \right) \!< 0\\ \ \ \ \ \ \ \ \ \ \ \ \ \ \ \ \ \ \ \ \ \ \ \ \ \  \wedge \  0 \!<\! \sqrt {\Delta {x^2} + {\rho ^2} + 2\rho \Delta x\cos \phi }  \!<\! D\\
\!\!\!0,\ \ \ {\rm{others}}
\end{array} \right.\!\!\!\!,
\end{aligned}
\end{footnotesize}
\end{equation}
where ${\varepsilon _i} = \arccos \left[ {\left( {{x_{{\rm{mid}}}} - {x_i}} \right)\sin \theta /D} \right]$.
\end{lemma}
\begin{IEEEproof}
See Appendix A.
\end{IEEEproof}

\begin{lemma}
The state vector of the IRS connection at the $i$-th measurement is given by
\begin{equation}
\setlength\abovedisplayskip{3pt}
\setlength\belowdisplayskip{3pt}
\begin{aligned}
{{\bf{S}}^{{\mathcal{I}}^k}}\!\!\left( i \right) &= {{\bf{S}}^{{\mathcal{I}}^k}}\!\!\left( 0 \right) \cdot {{\bf{T}}^{{\mathcal{I}}^k}}\!\!\left( 0 \right){{\bf{T}}^{{\mathcal{I}}^k}}\!\!\left( 1 \right)  \cdots  {{\bf{T}}^{{\mathcal{I}}^k}}\!\!\left( {i - 1} \right)\\
 &= \left[ {s_{1}^{{\mathcal{I}}^k}\left( i \right),s_{2}^{{\mathcal{I}}^k}\left( i \right),s_{3}^{{\mathcal{I}}^k}\left( i \right),s_{4}^{{\mathcal{I}}^k}\left( i \right)} \right],
\end{aligned}
\end{equation}
where $s_{m}^{{\mathcal{I}}^k}\left( i \right), k \in \left\{ {o, t} \right\},m \in \left\{ {1,2,3,4} \right\}$ represents the probability that the user is in the IRS connection state of ${\mathcal I}^k_{m}$ at the $i$-th measurement, ${{\bf{S}}^{{\mathcal{I}}^k}}\!\!\left( 0 \right)$ is the initial state, and the elements of ${{\bf{S}}^{{\mathcal{I}}^o}}\!\!\left( 0 \right)$ and ${{\bf{S}}^{{\mathcal{I}}^t}}\!\!\left( 0 \right)$ are given by
\begin{equation}
\setlength\abovedisplayskip{3pt}
\setlength\belowdisplayskip{3pt}
\begin{small}
\begin{aligned}
\!\!\!\!\!\!\!s_{1}^{{{\mathcal{I}}^o}}\!\!\!\left( 0 \right)\! =\!\! \frac{{p_{1,1}^{{{\mathcal{I}}^o}}\left( 0 \right)p_{2,3}^{{{\mathcal{I}}^o}}\left( 0 \right)}}{{p_{2,3}^{{{\mathcal{I}}^o}}\left( 0 \right) \!+\! p_{1,2}^{{{\mathcal{I}}^o}}\left( 0 \right)}}, \! s_{2}^{{{\mathcal{I}}^o}}\!\!\left( 0 \right) \!\!=\!\! \frac{{p_{1,2}^{{{\mathcal{I}}^o}}\left( 0 \right)p_{2,3}^{{{\mathcal{I}}^o}}\left( 0 \right)}}{{p_{2,3}^{{{\mathcal{I}}^o}}\left( 0 \right) \!+\! p_{1,2}^{{{\mathcal{I}}^o}}\left( 0 \right)}},\\[1mm]
\!\!\!\!\!\!\!s_{3}^{{{\mathcal{I}}^o}}\!\!\left( 0 \right) \!=\!\! \frac{{p_{1,2}^{{{\mathcal{I}}^o}}\left( 0 \right)p_{2,3}^{{{\mathcal{I}}^o}}\left( 0 \right)}}{{p_{2,3}^{{{\mathcal{I}}^o}}\left( 0 \right) \!+\! p_{1,2}^{{{\mathcal{I}}^o}}\left( 0 \right)}}, \! s_{4}^{{{\mathcal{I}}^o}}\!\!\left( 0 \right) \!\!=\!\! \frac{{p_{1,2}^{{{\mathcal{I}}^o}}\left( 0 \right)p_{2,4}^{{{\mathcal{I}}^o}}\left( 0 \right)}}{{p_{2,3}^{{{\mathcal{I}}^o}}\left( 0 \right) \!+\! p_{1,2}^{{{\mathcal{I}}^o}}\left( 0 \right)}},
\end{aligned}
\end{small}\!\!\!\!\!\!\!\!
\end{equation}
\begin{equation}
\setlength\abovedisplayskip{0pt}
\setlength\belowdisplayskip{3pt}
\begin{small}
s_{1}^{{{\mathcal I}^t}}\!\!\left( 0 \right) = 1, \ s_{2}^{{{\mathcal I}^t}}\!\!\left( 0 \right) = s_{3}^{{{\mathcal I}^t}}\!\!\left( 0 \right) = s_{4}^{{{\mathcal I}^t}}\!\!\left( 0 \right) = 0 \ .
\end{small}
\end{equation}
\end{lemma}
\begin{IEEEproof}
As for ${{\bf{S}}^{{\mathcal{I}}^o}}\!\!\!\left( 0\right)$, since the user has been residing at the original BS for a while at the initial location, it can be assumed that the IRS connection state of the original BS is in its steady state. Thus, we have the following equation
\begin{equation}
\setlength\abovedisplayskip{3pt}
\setlength\belowdisplayskip{3pt}
\begin{small}
\begin{aligned}
\left\{ \begin{array}{l}
\sum\limits_{k \in \left\{ {1,2,3,4} \right\}} \!\!\!{s_k^{{{\mathcal I}^o}}\left( 0 \right)}  = 1\\
s_k^{{{\mathcal I}^o}}\!\!\left( 0 \right) =\!\!\!\!\!\!\! \sum\limits_{j \in \left\{ {1,2,3,4} \right\}} \!\! {s_j^{{{\mathcal I}^o}}\!\!\left( 0 \right)p_{j,k}^{{{\mathcal I}^o}}\left( 0 \right)} ,k \in \left\{ {1,2,3,4} \right\}
\end{array} \right.\!\!.
\end{aligned}
\end{small}
\end{equation}
By solving equations in (15), we obtain (13). 

As for ${{\bf{S}}^{{\mathcal{I}}^t}}\!\!\! \left(0 \right)$, since the user is not handed over to the target BS at the initial location, it cannot be connected to the IRS of target BS; thus, only $s_{1}^{{\mathcal I}^t}\left(0\right)$ takes 1.  
\end{IEEEproof}
\vspace{-0.4cm}
\subsection{Distance and Angle Distribution of the IRS}
To obtain the exact IRS channel gain and analyze the effect of the IRS on the HO process, distributions of the distance and angle between the typical user and the IRS needs to be derived. Additionally, the distributions varies under different IRS connection states, which need to be analyzed separately. Therefore, we have the following lemmas:

\begin{lemma}
The pdfs of the distance between the typical user and the serving IRS at $i$-th measurement under 
${\cal{I}}^{o}_2$, ${\cal{I}}^{o}_4$, ${\cal{I}}^{t}_2$, and ${\cal{I}}^{t}_4$ are given by 
\begin{equation}
\begin{aligned}
&f_{d}\!\left( d\! \left| {i,{\cal{I}}_2^o} \right. \!\right) = \left\{ \begin{array}{l}
\!\!\!\!\left( {\frac{{\partial S_{i,d}^s}}{{\partial d}} \!-\! \frac{{\partial S_{i,d}^{s, \cap }}}{{\partial d}}} \right)\frac{{{\lambda _r}{e^{ - {\lambda _r}\left( {S_{i,d}^s - S_{i,d}^{s, \cap }} \right)}}}}{{1 - {e^{ - {\lambda _r}\left( {S_i^s - S_i^{s, \cap }} \right)}}}},\\
\ \ \ \ \ \ \ D \!-\! \Delta x \!<\! d \!\le\! D  \wedge  0 \!\le\! {x_i} \!\le\! {x_{{\rm{mid}}}} \!+\! \frac{D}{{\sin \theta }}\\
\!\! 0, \ \ \ \ \ \rm{otherwise}
\end{array} \right.\!\!\!,\\
&f_{d}\!\left( d\! \left| {i,{\cal{I}}_4^o} \right. \!\right) = \left\{ \begin{array}{ll}
\!\!\!\! \frac{{\partial S_{i,d}^s}}{{\partial d}}\frac{{{\lambda _r}{e^{ - {\lambda _r}S_{i,d}^s}}}}{{1 - {e^{ - {\lambda _r}S_i^s}}}},& \!\!\! 0 \!<\! d\! \le \! D \wedge 0 \!\le\! {x_i} \!\le\! {x_{{\rm{mid}}}} \!+\! \frac{D}{{\sin \theta }}\\
\!\! 0, & \!\!\!\rm{otherwise}
\end{array} \right.\!\!\!\!\!,\\
& f_{d}\!\left( d\! \left| {i,{\cal{I}}_2^t} \right. \!\right)  = \left\{ \begin{array}{l}
\!\!\! \left( {\frac{{\partial S_{i,d}^t}}{{\partial d}} \!-\! \frac{{\partial S_{i,d}^{t, \cap }}}{{\partial d}}} \right)\frac{{{\lambda _r}{e^{ - {\lambda _r}\left( {S_{i,d}^t - S_{i,d}^{t, \cap }} \right)}}}}{{1 - {e^{ - {\lambda _r}\left( {S_i^t - S_i^{t, \cap }} \right)}}}},\\
\ \ \ \ \ \ \ \ D \!-\! \Delta x \!<\! d \!\le\! D \wedge {x_{{\rm{mid}}}} \!-\! \frac{D}{{\sin \theta }} \!\le\! {x_i}\\
\!\! 0, \ \ \ \ \ \ \rm{otherwise}
\end{array} \right.\!\!\!,\\
& f_{d}\!\left( d\! \left| {i,{\cal{I}}_4^t} \right. \!\right) = \left\{ \begin{array}{ll}
\!\!\! \frac{{\partial S_{i,d}^t}}{{\partial d}}\frac{{{\lambda _r}{e^{ - {\lambda _r}S_{i,d}^t}}}}{{1 - {e^{ - {\lambda _r}S_i^t}}}},&\!\!\! 0 \!<\! d\! \le \! D \wedge  \! {x_{{\rm{mid}}}} \!+\! \frac{D}{{\sin \theta }}\! \le \! {x_i} \\
0,&\!\!\!{\rm{otherwise}}
\end{array} \right.\!\!\!,
\end{aligned}
\end{equation}
where $S^o_{i,d}$ ($S^t_{i,d}$) is the area of the region where the IRS of original (target) BS may exist with a distance from the user less than $d$ at the $i$-th measurement and is given by 
\begin{equation}
\begin{small}
\begin{aligned}
& S_{i,d}^o = \left\{ \begin{array}{l}
\!\! \pi {d^2},\ \ \ \ \ \ \ \ \ \ \ \ \ \ \ \ \ \ \ \ \ \ \ 0 \!<\! d \!\le\! D \wedge 0 \!\le {x_i} \le {x_{{\rm{mid}}}} \!-\! \frac{d}{{\sin \theta }}\\
\!\! \pi\! {d^2} \!-\! \frac{{{d^2}\left( {\mu  - \sin \mu } \right)}}{2},\\
\ \ \ \left| {{x_{{\rm{mid}}}} \!-\! {x_i}} \right|\!\sin \theta  \!<\! d \!\le\! D \wedge {x_{{\rm{mid}}}} \!-\! \frac{D}{{\sin \theta }} \!<\! {x_i} \!\le\! {x_{{\rm{mid}}}} \!+\! \frac{D}{{\sin \theta }}
\end{array} \right.,\\
& S_{i,d}^t = \left\{ \begin{array}{l}
\!\!\frac{{{d^2}\left( {\mu  - \sin \mu } \right)}}{2},\\
\ \ \ \ \!\!\left| {{x_i} \!-\! {x_{{\rm{mid}}}}} \right|\sin \theta  \!< \!d \!\le\! D\wedge{x_{{\rm{mid}}}} \!-\! \frac{D}{{\sin \theta }} \!<\! {x_i} \!\le\! {x_{{\rm{mid}}}} \!+\! \frac{D}{{\sin \theta }}\\
\!\!\pi {d^2},\ \ \ \ \ \ \ \ \ \ \ \ \ \ \ \ \ \ \ \ \ \ \ 0 < d \le D \wedge {x_{{\rm{mid}}}} - \frac{d}{{\sin \theta }} \le {x_i}
\end{array} \right.,
\end{aligned}
\end{small}
\end{equation}
where $\mu  = 2\arccos \frac{{\left| {{x_{{\rm{mid}}}} - {x_i}} \right|\sin \theta }}{d}$, $S^{o,\cap}_{i,d}$ ($S^{t,\cap}_{i,d}$) is the area of overlap of regions corresponding to $S^{o}_{i-1}$ and $S^{o}_{i,d}$ ($S^{t}_{i-1}$ and $S^{t}_{i,d}$) and is given by
\begin{equation}
\begin{footnotesize}
\begin{aligned}
& S_{i,d}^{o, \cap } = \left\{ \begin{array}{l}
{d^2}\arccos \! \left( {\frac{{{d^2} + \Delta {x^2} - {D^2}}}{{2\Delta xd}}} \right) + {D^2}\arccos \! \left( {\frac{{{D^2} + \Delta {x^2} - {d^2}}}{{2\Delta xD}}} \right)\\
 \ \ \ \ - \frac{1}{2}\sqrt {\left( {d \!+\! \Delta x \!+\! D} \right)\left( {d \!+\! \Delta x \!-\! D} \right)\left( {D \!+\! d \!-\! \Delta x} \right)\left( {D \!+\! \Delta x \!-\! d} \right)} ,\\
\ \ \ \ \ \ \ \ \ \ \ \ \ \ \ \ \ \ \ \ \ \ \ \ \ \ \ \ \ \ \ \ \ \ \ \ \ \ \ \ \ \ \ \  0 \!\le\! x \!\le\! {x_{{\rm{mid}}}} \!-\! \frac{D}{{\sin \theta }} \!+\! \Delta x\\
\!\!\! \int\limits_0^d \! \int\limits_0^{2\pi }\!\! {\Delta S_i^{o, \cap }\!\left( {\phi ,\rho } \right)\rho {\rm{d}}\phi {\rm{d}}\rho } ,\ \ \ {x_{{\rm{mid}}}} \!-\! \frac{D}{{\sin \theta }} \!+\! \Delta x \!<\! {x_i} \!\le\! {x_{{\rm{mid}}}} \!+\! \frac{D}{{\sin \theta }}
\end{array} \right.\!\!\!,\\
&S_{i,d}^{t, \cap } = \left\{ \begin{array}{l}
\!\!\! \int\limits_0^d \! \int\limits_0^{2\pi }\!\! {\Delta S_i^{t, \cap }\!\left( {\phi ,\rho } \right)\rho {\rm{d}}\phi {\rm{d}}\rho } ,\ \ \ {x_{{\rm{mid}}}} \!-\! \frac{D}{{\sin \theta }} \!+\! \Delta x \!<\! {x_i} \!\le\! {x_{{\rm{mid}}}} \!+\! \frac{D}{{\sin \theta }}\\
{d^2}\arccos \! \left( {\frac{{{d^2} + \Delta {x^2} - {D^2}}}{{2\Delta xd}}} \right) + {D^2}\arccos \! \left( {\frac{{{D^2} + \Delta {x^2} - {d^2}}}{{2\Delta xD}}} \right)\\
 \ \ \ \ - \frac{1}{2}\sqrt {\left( {d \!+\! \Delta x \!+\! D} \right)\left( {d \!+\! \Delta x \!-\! D} \right)\left( {D \!+\! d \!-\! \Delta x} \right)\left( {D \!+\! \Delta x \!-\! d} \right)} , \\
\ \ \ \ \ \ \ \ \ \ \ \ \ \ \ \ \ \ \ \ \ \ \ \ \ \ \ \ \ \ \ \ \ \ \ \ \ \ \ \ \ \ \ \ \ \ \ \ \ \ \ {x_{{\rm{mid}}}} + \frac{D}{{\sin \theta }} \le {x_i}
\end{array} \right.\!\!\!\!\!\!,
\end{aligned}
\end{footnotesize}
\end{equation}
where $\Delta S_i^{k, \cap } \! {\left( {\phi ,\rho } \right)}, k \in \{o,t\}$ is given in (11), $\frac{{\partial S_{i,d}^{k}}}{{\partial d}}$ ($\frac{{\partial S_{i,d}^{k, \cap}}}{{\partial d}}$) is the derivative of $S_{i,d}^{k}$ ($S_{i,d}^{k, \cap}$) with respect to $d$ and is given by  
\begin{equation}
\begin{footnotesize}
\begin{aligned}
&\frac{{\partial S_{i,d}^o}}{{\partial d}} = \left\{ \begin{array}{l}
\!\!\! 2\pi d, \ \ \ \ \ \ \ \ \ \ \ \ \ \ \ \ \ \ \ \ \ \ \ \ \ \ \ \ 0 \!<\! d \!\le\! D \wedge 0\! \le\! {x_i} \!\le\! {x_{{\rm{mid}}}} \!-\! \frac{d}{{\sin \theta }}\\
\!\!\! \left( {2\pi  \!-\! \mu  \!+\! \sin \mu  \!-\! \frac{{\mu 'd}}{2}\left( {1 \!-\! \cos \mu } \right)} \right)\!d,\\
\ \ \ \ \ \  \left| {{x_{{\rm{mid}}}} \!-\! {x_i}} \right|\sin \theta  \!<\! d \!\le\! D  \wedge {x_{{\rm{mid}}}} \!-\! \frac{D}{{\sin \theta }} \!<\! {x_i} \!\le\! {x_{{\rm{mid}}}}\! +\! \frac{D}{{\sin \theta }}
\end{array} \right.,
\end{aligned}
\end{footnotesize}\notag
\end{equation}
\begin{equation}
\begin{footnotesize}
\begin{aligned}
& \frac{{\partial S_{i,d}^t}}{{\partial d}} = \left\{ \begin{array}{l}
\!\!\! \left( {\mu  \!-\! \sin \mu  \!+\! \frac{{\mu 'd}}{2}\left( {1 \!-\! \cos \mu } \right)} \right)\!d,\\
\ \ \ \ \ \  \left| {{x_{{\rm{mid}}}} \!-\! {x_i}} \right|\sin \theta  \!<\! d \!\le\! D  \wedge {x_{{\rm{mid}}}} \!-\! \frac{D}{{\sin \theta }} \!<\! {x_i} \!\le\! {x_{{\rm{mid}}}}\! +\! \frac{D}{{\sin \theta }}\\
\!\!\! 2\pi d,\ \ \ \ \ \ \ \ \ \ \ \ \ \ \ \ \ \ \ \ \ \ \ \ \ \ \ \ \ \ \ \ \ 0 \!<\! d \!\le\! D \wedge {x_{{\rm{mid}}}} \!-\! \frac{d}{{\sin \theta }} \!\le\! {x_i}
\end{array} \right.,\\
&\frac{{\partial S_{i,d}^{o, \cap }}}{{\partial d}} = \left\{ \begin{array}{ll}
\!\!\! 2d\!\arccos\! \left( {\frac{{{d^2} - {D^2} + \Delta {x^2}}}{{2d\Delta x}}} \right),&0\! \le\! x \!\le\! {x_{{\rm{mid}}}} \!-\! \frac{D}{{\sin \theta }} \!+\! \Delta x\\
\!\!\! \int\limits_0^{2\pi } {\Delta S_i^{o, \cap }\left( {\phi ,d} \right)d{\rm{d}}\phi } ,&{x_{{\rm{mid}}}} \!-\! \frac{D}{{\sin \theta }} \!+\! \Delta x \!<\! {x_i} \!\le\! {x_{{\rm{mid}}}} \!+\! \frac{D}{{\sin \theta }}
\end{array} \right.,\\
&\frac{{\partial S_{i,d}^{t, \cap }}}{{\partial d}} = \left\{ \begin{array}{ll}
\!\!\!\int\limits_0^{2\pi } \!{\Delta S_i^{t, \cap }\!\left( {\phi ,d} \right)\!d{\rm{d}}\phi } ,&{x_{{\rm{mid}}}} \!-\! \frac{D}{{\sin \theta }} \!+\! \Delta x \!<\! {x_i}\!\le\! {x_{{\rm{mid}}}} \!+\! \frac{D}{{\sin \theta }}\\
\!\!\!2d\arccos \left( {\frac{{{d^2} - {D^2} + \Delta {x^2}}}{{2d\Delta x}}} \right),&{x_{{\rm{mid}}}} + \frac{D}{{\sin \theta }} \le {x_i}
\end{array} \right.,
\end{aligned}
\end{footnotesize}
\end{equation}
where $\mu ' =  \frac{{2\left| {{x_{{\rm{mid}}}} - {x_i}} \right|\sin \theta }}{{d\sqrt {{d^2} - {{\left( {{x_{{\rm{mid}}}} - {x_i}} \right)}^2}{{\sin }^2}\theta } }}$.
\end{lemma}
\begin{IEEEproof}
See Appendix B.
\end{IEEEproof}
\begin{lemma}
The pdfs of the angle from the user trajectory to the serving IRS at $i$-th measurement under ${\cal{I}}^{o}_2$, ${\cal{I}}^{o}_4$, ${\cal{I}}^{t}_2$, and ${\cal{I}}^{t}_4$ are given by 
\begin{equation}
\begin{small}
\begin{aligned}
& {f_{\varphi '}}\!\!\left( {\varphi '\left| {i,d,{\cal{I}}_2^o} \right.} \right) = \\
& \left\{ \begin{array}{ll}
\!\!\! \frac{1}{\kappa },&\varphi ' \!\!\in\!\! \left( {0,\frac{\kappa }{2}} \right) \!\cup\! \left( {2\pi  \!-\! \frac{\kappa }{2},2\pi } \right] \wedge 0 \!\le\! {x_i} \!\le\! {x_{{\rm{mid}}}} \!-\! \frac{D}{{\sin \theta }}\\
\!\!\! \frac{1}{{2\pi  - \mu }},&\varphi ' \!\!\in\!\! \left( {\frac{{\mu  + \pi }}{2} \!-\! \theta ,\frac{{5\pi  - \mu }}{2} \!-\! \theta } \right)\\
& \wedge \left| {{x_i} \!-\! {x_{{\rm{mid}}}}} \right|\sin \theta  \!<\! d \!\le\! D \wedge{x_{{\rm{mid}}}} \!+\! \frac{D}{{\sin \theta }} \!\le\! {x_i} \!\le \!{x_{{\rm{mid}}}} \!-\! \frac{D}{{\sin \theta }}
\end{array} \right.\!\!\!,\\
& {f_{\varphi '}}\!\!\left( {\varphi '\left| {i,d,{\cal{I}}_4^o} \right.} \right) = \\
&\left\{ \begin{array}{ll}
\!\!\!\frac{1}{{2\pi }},&\varphi ' \!\in\! \left( {0,2\pi } \right] \wedge 0 \le {x_i} \le {x_{{\rm{mid}}}} - \frac{d}{{\sin \theta }}\\
\!\!\!\frac{1}{{2\pi  - \mu }},&\varphi ' \in \left( {\frac{{\mu  + \pi }}{2} - \theta ,\frac{{5\pi  - \mu }}{2} - \theta } \right)\\
&\wedge \left| {{x_{{\rm{mid}}}} - {x_i}} \right|\sin \theta  \!<\!\! d \!\le\!D \wedge {x_{{\rm{mid}}}} \!-\! \frac{D}{{\sin \theta }} \!<\! {x_i} \!\le\! {x_{{\rm{mid}}}} \!+\! \frac{D}{{\sin \theta }}
\end{array} \right.\!\!\!,\\
&{f_{\varphi '}}\!\!\left( {\varphi '\left| {i,d,{\cal{I}}_2^t} \right.} \right) = \\ &\left\{ \begin{array}{ll}
\!\!\! \frac{1}{\mu },&\varphi ' \!\in\! \left( {0,\frac{{\mu  + \pi }}{2} \!-\! \theta } \right) \!\cup\! \left( {\frac{{5\pi  - \mu }}{2} \!-\! \theta ,2\pi } \right]\\
&\wedge \left| {{x_{{\rm{mid}}}} \!-\! {x_i}} \right|\sin \theta  \!<\! d \!\le\! D \wedge {x_{{\rm{mid}}}} \!- \! \frac{D}{{\sin \theta }} \!<\! {x_i} \!\le\! {x_{{\rm{mid}}}} \!+\! \frac{D}{{\sin \theta }}\\
\!\!\! \frac{1}{\kappa },&\varphi ' \in \left( {0,\frac{\kappa }{2}} \right) \cup {\left( {2\pi  - \frac{\kappa }{2},2\pi } \right]} \wedge {x_{{\rm{mid}}}} + \frac{D}{{\sin \theta }} \le {x_i}
\end{array} \right.\!\!\!,\\
& {f_{\varphi '}}\!\!\left( {\varphi '\left| {i,d,{\cal{I}}_4^t} \right.} \right) = \\
& \left\{ \begin{array}{ll}
\!\!\! \frac{1}{\mu },&\varphi ' \!\in\! \left( {0,\frac{{\mu  + \pi }}{2} \!-\! \theta } \right) \!\cup\! \left( {\frac{{5\pi  - \mu }}{2} \!-\! \theta ,2\pi } \right]\\
 &\wedge \left| {{x_{{\rm{mid}}}} \!-\! {x_i}} \right|\sin \theta  \!<\! d \!\le\! D \wedge {x_{{\rm{mid}}}} \!-\! \frac{D}{{\sin \theta }} \!<\! {x_i} \!\le\! {x_{{\rm{mid}}}} \!+\! \frac{D}{{\sin \theta }}\\
\!\!\! \frac{1}{{2\pi }},&{x_{{\rm{mid}}}} \!+\! \frac{D}{{\sin \theta }} \!<\! {x_i}
\end{array} \right.\!,
\end{aligned}
\end{small}
\end{equation}
where $\kappa  = 2\arccos \frac{{{D^2} - \Delta {x^2} - {d^2}}}{{2\Delta xd}}$ and $\mu  = 2\arccos \frac{{\left| {{x_{{\rm{mid}}}} - {x_i}} \right|\sin \theta }}{d}$. 
\end{lemma}
\begin{IEEEproof}
See Appendix C.
\end{IEEEproof}

To consider the IRS random scattering when there is no IRS serving, the following lemma is given.
\begin{lemma}
For states without IRS serving (i.e., ${\cal{I}}^{o}_1$, ${\cal{I}}^{o}_3$, ${\cal{I}}^{t}_1$, and ${\cal{I}}^{t}_3$), the pdfs of the distance from the user to the nearest IRS and the angle from the user trajectory to it at $i$-th measurement are given by
\begin{equation}
\setlength\abovedisplayskip{3pt}
\setlength\belowdisplayskip{3pt}
\begin{small}
\begin{aligned}
&f_{d}\!\left( d\! \left| {i,{\cal{I}}_m^k} \right. \!\right) = 2\pi d{e^{ - \pi \left( {{d^2} - {D^2}} \right)}},d > D,\\
&{f_{\varphi '}}\!\!\left( {\varphi '\left| {i,d,{\cal{I}}_m^k} \right.} \right) = \frac{1}{{2\pi }},\varphi ' \in \left( {0,2\pi } \right],
\end{aligned}
\end{small}
\end{equation}
where $k \in \left\{ {o,t} \right\}$ and $m \in \left\{ {1,3} \right\}$.
\end{lemma}
\vspace{-0.4cm}
\subsection{Handover Analysis}
As discussed in Section II.C, the HO process involves three steps: HO triggering, TTT timing, and HO execution. As shown in Fig. 3(b), the HO process is modeled as a discrete-time model, whose state space contains all steps of the HO process. The states of the HO process are denoted by ${\mathcal{H}}_0$, ${\mathcal{H}}_1$, $\cdots$, ${\mathcal{H}}_j$, and ${\mathcal{H}}_c$.
The definitions of the HO states are as follows: ${\mathcal{H}}_0$
denotes the state in which A3 event remains untriggered. ${\mathcal{H}}_1$ indicates that HO is just triggered and the TTT timer starts. If the measurement results satisfy the HO condition after HO triggering, the state of the typical user transitions from ${\mathcal{H}}_q$ to ${\mathcal{H}}_{q+1}$ $\left( {q \in \left\{ {1, \cdots ,j - 1} \right\}} \right)$ until the TTT timer expires, thus the number of states in the discrete-time model of HO process is determined by the settings of TTT and the measurement period, where $j \!=\! \left\lfloor {\frac{{{T_t}}}{{{T_d}}}} \right\rfloor \!+\! 1$, and $\left\lfloor \cdot \right\rfloor $ is the floor function. ${\mathcal{H}}_{j}$ indicates that the HO condition is satisfied during the entire TTT and the HO is executed. ${\mathcal{H}}_{c}$ denotes that HO is complete.

Based on the definitions of the states of the HO process, the corresponding state vector and state transition matrix of the discrete-time model of the HO process are constructed as follows:

\begin{lemma}
The state transition matrix of the discrete-time model of the HO process in IRS-assisted networks is given by
\begin{equation}
\setlength\abovedisplayskip{3pt}
\setlength\belowdisplayskip{3pt}
\begin{small}
{{\rm{\bf{T}}}^{\mathcal{H}}}\!\!\left( i \right) \!=\! \left[ {\begin{array}{*{20}{c}}
{1 - {p^{\cal H}}\left( i \right)}&{{p^{\cal H}}\left( i \right)}&0& \cdots &0&0\\
{1 - {p^{\cal H}}\left( i \right)}&0&{{p^{\cal H}}\left( i \right)}& \cdots &0&0\\
 \vdots & \vdots & \vdots & \ddots & \vdots & \vdots \\
{1 - {p^{\cal H}}\left( i \right)}&0&0& \cdots &{{p^{\cal H}}\left( i \right)}&0\\
0&0&0& \cdots &0&1\\
0&0&0& \cdots &0&1
\end{array}} \right],
\end{small}
\end{equation}
where its size is $\left( {j + 2} \right) \!\times\! \left( {j + 2} \right)$, $j \!=\! \left\lfloor {\frac{{{T_t}}}{{{T_d}}}} \right\rfloor  \!+\! 1$, the element of the $n$-th row and $m$-th column represents the probability that the $n$-th state transitions  to the $m$-th state, the order of states of HO process is ${\mathcal{H}}_0$, ${\mathcal{H}}_1$, $\cdots$, ${\mathcal{H}}_j$, ${\mathcal{H}}_c$, and ${p^{\mathcal H}}\!\left( i \right)$ is the probability of the typical user meeting the HO trigger condition (i.e., the condition of A3 event) at $i$-th measurement and its expression is integrated concerning all IRS connection states, which is given by
\begin{equation}
\setlength\abovedisplayskip{3pt}
\setlength\belowdisplayskip{3pt}
\begin{small}
\begin{aligned}
{p^{\cal H}}\left( i \right) = \sum\limits_{m \in \left\{ {1,2,3,4} \right\}} {s_m^{{{\cal I}^o}}\!\!(i){\mathbb P}\left( {{\rm{HO}}\left| {i,{\cal I}_m^o} \right.} \right)} ,
\end{aligned}
\end{small}
\end{equation}
where 
\begin{equation}
\setlength\abovedisplayskip{3pt}
\setlength\belowdisplayskip{3pt}
\begin{small}
\begin{aligned}
&{\mathbb P}\left( {{\rm{HO}}\left| {i,{\cal I}_m^o} \right.} \right) = {{\mathbb E}_{d,\varphi '}}\left[ {{\mathbb P}\left( {{\rm{HO}}\left| {i,d,\varphi ',{\cal I}_m^o} \right.} \right)} \right],\\
&{\mathbb P}\left( {{\rm{HO}}\left| {i,d,\varphi ',{\cal I}_m^o} \right.} \right) \!=\! \left\{ {\begin{array}{ll}
1,&{\eta _{{\rm{HO}},m}}\left( {{x_i},d,\varphi '} \right) \ge {\gamma _{{\rm{HO}}}}\\
0,&{\rm{otherwise}}
\end{array}} \right.,
\\
&{\eta _{{\rm{HO}},m}}\left( {{x_i},d,\varphi '} \right) \!=\! \left\{ \begin{array}{ll}
\!\!\!{{{\Gamma _{sc}}\left( {{x_t},d,{\varphi _t}} \right)} \mathord{\left/
 {\vphantom {{{\Gamma _{sc}}\left( {{x_t},d,{\varphi _t}} \right)} {{\Gamma _{sc}}\left( {{x_o},d,{\varphi _o}} \right)}}} \right.
 \kern-\nulldelimiterspace} {{\Gamma _{sc}}\left( {{x_o},d,{\varphi _o}} \right)}},&\!\!\!m \in \left\{ {1,3} \right\}\\
\!\!\!{{{\Gamma _{sc}}\left( {{x_t},d,{\varphi _t}} \right)} \mathord{\left/
 {\vphantom {{{\Gamma _{sc}}\left( {{x_t},d,{\varphi _t}} \right)} {{\Gamma _{sc}}\left( {{x_o},d,{\varphi _o}} \right)}}} \right.
 \kern-\nulldelimiterspace} {{\Gamma _{bf}}\left( {{x_o},d,{\varphi _o}} \right)}},&\!\!\!m \in \left\{ {2,4} \right\}
\end{array} \right.\!\!\!,
\end{aligned}
\end{small}
\end{equation}
where pdfs of $d$ and $\varphi '$ are given in \emph{Lemma 3, Lemma 4, and Lemma 5}.
\end{lemma}
\begin{IEEEproof}
According to the definition of the states of the HO process, if the HO condition is satisfied within the TTT duration, the state transitions from ${\mathcal{H}}_0$ to ${\mathcal{H}}_1$, $\cdots$, ${\mathcal{H}}_{j}$ in turn. Otherwise, the state returns to ${\mathcal{H}}_0$, which is the state in which HO is not triggered.

The conditional probabilities of satisfying the HO condition are obtained from (4) and the pdfs of the user-IRS distance and angle given in \emph{Lemma 3, Lemma 4, and Lemma 5}. By integrating the conditional probabilities and state vector in \emph{Lemma 2}, (23) is derived.
\end{IEEEproof}

\begin{lemma}
The state vector of the HO process in IRS-assisted networks at the $i$-th measurement is given by
\begin{equation}
\setlength\abovedisplayskip{3pt}
\setlength\belowdisplayskip{3pt}
\begin{aligned}
&{{\bf{S}}^{\mathcal{H}}}\!\!\left( i \right) = {{\bf{S}}^{\mathcal{H}}}\!\!\left( 0 \right) {{\rm{\bf{T}}}^{\mathcal{H}}}\!\left( 0 \right){{\rm{\bf{T}}}^{\mathcal{H}}}\!\left( 1 \right) \cdots {{\rm{\bf{T}}}^{\mathcal{H}}}\!\left( {i - 1} \right)\\
 &=\! \left[ {s_0^{\mathcal{H}}\!\left( i \right),s_1^{\mathcal{H}}\!\left( i \right),s_2^{\mathcal{H}}\!\left( i \right), \cdots \! ,s_{j - 1}^{\mathcal{H}}\!\left( i \right),s_j^{\mathcal{H}}\!\left( i \right),s_c^{\mathcal{H}}\!\left( i \right)} \right]\!,\\[1mm]
\end{aligned}
\end{equation}
where ${{\bf{S}}^{\mathcal{H}}}\!\!\left( 0 \right) = \left[ {1,0,0, \cdots ,0} \right]$.
\end{lemma}

For the HO process, we are concerned with the effect of IRS implementation on locations where the HO is triggered and locations where the HO is executed; therefore, we give the following theorem and corollary.

\begin{theorem}
At $i$-th measurement, the probability that the typical user triggers an HO and the probability that the typical user executes an HO in IRS-assisted networks are given by
\begin{equation}
\setlength\abovedisplayskip{3pt}
\setlength\belowdisplayskip{3pt}
\begin{aligned}
&{{\mathbb{P}}_{ht}}\!\left( {{x_i}} \right) = {\Xi _2}\left[ {{{\bf{S}}^{\mathcal{H}}}\!\!\left( 0 \right){{\rm{\bf{T}}}^{\mathcal{H}}}\!\left( 0 \right){{\rm{\bf{T}}}^{\mathcal{H}}}\!\left( 1 \right) \cdots {{\rm{\bf{T}}}^{\mathcal{H}}}\!\left( {i - 1} \right)} \right] = s_1^{\mathcal{H}}\!\left( i \right),\\
&{{\mathbb{P}}_{ho}}\!\left( {{x_i}} \right) = {\Xi _{j+1}}\left[ {{{\bf{S}}^{\mathcal{H}}}\!\!\left( 0 \right){{\rm{\bf{T}}}^{\mathcal{H}}}\!\left( 0 \right){{\rm{\bf{T}}}^{\mathcal{H}}}\!\left( 1 \right) \cdots {{\rm{\bf{T}}}^{\mathcal{H}}}\!\left( {i - 1} \right)} \right] = s_j^{\mathcal{H}}\!\left( i \right),
\end{aligned}
\end{equation}
where ${\Xi _k}\left[ \cdot  \right]$ denotes the $k$-th element of the vector.
\end{theorem}

\begin{corollary}
The average distances from the initial location to the location where the HO is triggered and the location where the HO is executed in IRS-assisted networks are given by
\begin{equation}
\setlength\abovedisplayskip{3pt}
\setlength\belowdisplayskip{3pt}
\begin{aligned}
{\mathbb{E}}\left[ {{x^{ht}}} \right] \!=\! \sum\limits_{i = 1}^I {{{\mathbb{P}}_{ht}}\left( {{x_i}} \right) \!\cdot\! {x_i}} ,\ \ {\mathbb{E}}\left[ {{x^{ho}}} \right] \!=\! \sum\limits_{i = 1}^I {{{\mathbb{P}}_{ho}}\left( {{x_i}} \right) \!\cdot\! {x_i}} .
\end{aligned}
\end{equation}
\end{corollary}

\vspace{-0.3cm}
\subsection{Handover Failure Analysis}

Based on (5), an HOF occurs if an HO is triggered and the SIR drops below the threshold during TTT duration.
As shown in Fig. 3(c), the occurrence of HOF is modeled as a discrete-time model, where the HO condition and SIR threshold of HOF are considered simultaneously. States of HOF are denoted as  ${\mathcal{F}}_0$, ${\mathcal{F}}_1$, $\cdots$, ${\mathcal{F}}_{j-1}$, ${\mathcal{F}}_j$, ${\mathcal{F}}_t$, and ${\mathcal{F}}_c$.
The states of HOF are defined as follows: ${\mathcal{F}}_0$
denotes the state in which the A3 event remains untriggered. ${\mathcal{F}}_1$ indicates that HO is just triggered. If the measurement results satisfy the HO condition after the HO is triggered, and the HOF condition is not satisfied, the typical user transitions from ${\mathcal{F}}_q$ to ${\mathcal{F}}_{q+1}$ $\left( {q \in \left\{ {1, \cdots ,j - 1} \right\}}, j \!=\! \left\lfloor {\frac{{{T_t}}}{{{T_d}}}} \right\rfloor \!+\! 1 \right)$ until the TTT timer expires. ${\mathcal{F}}_{j}$ indicates that the HO condition is satisfied during the entire TTT and that the HOF is avoided. ${\mathcal{F}}_{t}$ denotes that the HOF condition is just satisfied, and ${\mathcal{F}}_{c}$ denotes that HOF occurs.

Based on the definition of HOF states, the corresponding 
state transition matrix and state vector of the discrete-time model of HOF are constructed as follows:
\begin{lemma}
The state transition matrix of the discrete-time model of HOF in IRS-assisted networks is given by
\begin{small}
\begin{equation}
\setlength\abovedisplayskip{3pt}
\setlength\belowdisplayskip{3pt}
\begin{aligned}
&{{\rm{\bf{T}}}^{\mathcal{F}}}\!\left( i \right) = \\[-1mm]&\left[ {\begin{array}{*{20}{c}}
\!\!{1 \!\!-\! {p^{\mathcal H}}\!\left( i \right)}\!&\!\!{{p^{\mathcal H}}\!\left( i \right)}\!\!&0& \!\cdots\! &0&0&\!\!0\!\!\\
\!\!{1 \!\!-\! {p^{\mathcal H}}\!\left( i \right)}\!&0&\!\!{{p^{\mathcal H}}\!\left( i \right) \!-\! {p^{\mathcal F}}\!\!\left( i \right)}\!\!& \!\!\cdots\!\! &0&\!\!{{p^{\mathcal F}}\!\!\left( i \right)}\!\!&\!\!0\!\!\\
 \vdots & \vdots & \vdots & \!\ddots\! & \vdots & \vdots & \!\!\vdots \!\!\\
\!\!{1 \!\!-\! {p^{\mathcal H}}\!\left( i \right)}\!&0&0&\!\cdots\!&\!\!{{p^{\mathcal H}}\!\!\left( i \right) \!-\! {p^{\mathcal F}}\!\left( i \right)}\!\!&\!{p^{\mathcal F}}\!\!\left( i \right)\!&\!\!0\!\!
\\
0&0&0& \!\cdots\! &1&0&\!\!0\!\!\\
0&0&0& \!\cdots\! &0&0&\!\!1\!\!\\
0&0&0& \!\cdots\! &0&0&\!\!1\!\!
\end{array}} \right]
\end{aligned},
\end{equation}
\end{small}

\noindent where its size is $\left( {j + 3} \right) \!\times\! \left( {j + 3} \right)$, the element of the $n$-th row and $m$-th column represents the probability that the $n$-th state transitions to the $m$-th state, the order of states of HOF is ${\mathcal{F}}_0$, ${\mathcal{F}}_1$, $\cdots$, ${\mathcal{F}}_{j-1}$, ${\mathcal{F}}_j$, ${\mathcal{F}}_t$, ${\mathcal{F}}_c$, and ${p^{\mathcal F}}\!\left( i \right)$ indicates the probability of the typical user meeting the HOF condition at the $i$-th measurement, and its expression is integrated concerning all IRS connection states, which is given by
\begin{equation}
\setlength\abovedisplayskip{3pt}
\setlength\belowdisplayskip{3pt}
\begin{small}
\begin{aligned}
{p^{\cal F}}\left( i \right) = \sum\limits_{m \in \left\{ {1,2,3,4} \right\}} {s_m^{{{\cal I}^o}}\!\!(i){\mathbb P}\left( {{\rm{HOF}}\left| {i,{\cal I}_m^o} \right.} \right)} ,
\end{aligned}
\end{small}
\end{equation}
where 
\begin{equation}
\setlength\abovedisplayskip{2pt}
\setlength\belowdisplayskip{2pt}
\begin{small}
\begin{aligned}
&{\mathbb P}\left( {{\rm{HOF}}\left| {i,{\cal I}_m^o} \right.} \right) = {{\mathbb E}_{d,\varphi '}}\left[ {{\mathbb P}\left( {{\rm{HOF}}\left| {i,d,\varphi ',{\cal I}_m^o} \right.} \right)} \right],\\
&{\mathbb P}\left( {{\rm{HOF}}\left| {i,d,\varphi ',{\cal I}_m^o} \right.} \right) = \left\{ {\begin{array}{*{20}{ll}}
\!\!\! 1,&{\eta _{{\rm{HOF}},m}}\left( {{x_i},d,\varphi '} \right) < {Q_{out}}\\
\!\!\! 0,&{\rm{otherwise}}
\end{array}} \right.,\\
&{\eta _{{\rm{HOF}},m}}\left( {{x_i},d,\varphi '} \right) \!=\! \left\{ \begin{array}{ll}
\!\!\!{{{\Gamma _{sc}}\left( {{x_o},d,{\varphi _o}} \right)} \mathord{\left/
 {\vphantom {{{\Gamma _{sc}}\left( {{x_o},d,{\varphi _o}} \right)} {{\Gamma _{sc}}\left( {{x_t},d,{\varphi _t}} \right)}}} \right.
 \kern-\nulldelimiterspace} {{\Gamma _{sc}}\left( {{x_t},d,{\varphi _t}} \right)}},&m \in \left\{ {1,3} \right\}\\
\!\!\!{{{\Gamma _{bf}}\left( {{x_o},d,{\varphi _o}} \right)} \mathord{\left/
 {\vphantom {{{\Gamma _{bf}}\left( {{x_o},d,{\varphi _o}} \right)} {{\Gamma _{sc}}\left( {{x_t},d,{\varphi _t}} \right)}}} \right.
 \kern-\nulldelimiterspace} {{\Gamma _{sc}}\left( {{x_t},d,{\varphi _t}} \right)}},&m \in \left\{ {2,4} \right\}
\end{array} \right.\!\!\!\!,
\end{aligned}
\end{small}
\end{equation}
where pdfs of $d$ and $\varphi '$ are given in \emph{Lemma 3, Lemma 4, and Lemma 5}.
\end{lemma}
\begin{IEEEproof}
According to the definition of the states of HOF, if the HO condition is satisfied and the condition of HOF is not satisfied within the TTT duration, the state transitions from ${\mathcal{F}}_0$ to ${\mathcal{F}}_1$, $\cdots$, ${\mathcal{F}}_{j}$ in turn. If the condition of HOF is satisfied within the TTT duration, the state transitions to ${\mathcal{F}}_{t}$ and ${\mathcal{F}}_{c}$ in turn. If the HO condition is not satisfied during TTT, the state returns to ${\mathcal{F}}_0$ (i.e., the state in which HO is not triggered).

The conditional probabilities of HOF are obtained based on (5) and pdfs of the user-IRS distance and angle given in \emph{Lemma 3, Lemma 4, and Lemma 5}. By integrating the conditional probabilities and state vector in \emph{Lemma 2}, (29) is derived.
\end{IEEEproof}
\begin{lemma}
The state vector of HOF in IRS-assisted networks at the $i$-th measurement is given by
\begin{equation}
\setlength\abovedisplayskip{3pt}
\setlength\belowdisplayskip{3pt}
\begin{aligned}
&{{\bf{S}}^{\mathcal{F}}}\!\!\left( i \right) = {{\bf{S}}^{\mathcal{F}}}\!\!\left( 0 \right) {{\rm{\bf{T}}}^{\mathcal{F}}}\!\left( 0 \right){{\rm{\bf{T}}}^{\mathcal{F}}}\!\left( 1 \right) \cdots {{\rm{\bf{T}}}^{\mathcal{F}}}\!\left( {i - 1} \right)\\[-1mm]
 &=\! \left[ {s_0^{\mathcal{F}}\!\left( i \right),s_1^{\mathcal{F}}\!\left( i \right), \cdots \! ,s_{j - 1}^{\mathcal{F}}\!\left( i \right),s_j^{\mathcal{F}}\!\left( i \right),s_t^{\mathcal{F}}\!\left( i \right),s_c^{\mathcal{F}}\!\left( i \right)} \right]\!,\\[1mm]
\end{aligned}
\end{equation}
where ${{\bf{S}}^{\mathcal{F}}}\!\!\left( 0 \right) = \left[ {1,0,0, \cdots ,0} \right]$.
\end{lemma}

Based on the discrete-time model for HOF, the expression of the HOF probability is given by the following theorem.

\begin{theorem}
The HOF probability of IRS-assisted networks is given by
\begin{equation}
\setlength\abovedisplayskip{3pt}
\setlength\belowdisplayskip{3pt}
\begin{aligned}
{\mathbb{P}}{_{hof}} \!=\! {\Xi _{j + 3}}\left[ {{{\bf{S}}^{\mathcal F}}\!\!\left( 0 \right)\!{{\rm{\bf{T}}}^{\mathcal{F}}}\!\!\left( 0 \right)\!{{\rm{\bf{T}}}^{\mathcal{F}}}\!\!\left( 1 \right) \cdots {{\rm{\bf{T}}}^{\mathcal{F}}}\!\left( {I - 1} \right)} \right] = s_c^{\mathcal F}\!\left( I \right),
\end{aligned}
\end{equation}
where ${\Xi_{j+3}}\left[ \cdot  \right]$ represents the $(j\!+\!3)$-th element of the vector.
\end{theorem}

\begin{corollary}
The average distance from the initial location to the location where HOF occurs is given by
\begin{equation}
\setlength\abovedisplayskip{3pt}
\setlength\belowdisplayskip{3pt}
\begin{aligned}
{\mathbb{E}}\left[ {{x^{hof}}} \right] \!=\! {{\sum\limits_{i = 1}^I {s_t^{\mathcal F}\!\!\left( {{i}} \right) \cdot {x_i}} } \mathord{\left/
 {\vphantom {{\sum\limits_{i = 1}^I {s_t^F\left( {{x_i}} \right) \cdot {x_i}} } {{P_{hof}}}}} \right.
 \kern-\nulldelimiterspace} {{{\mathbb P}_{hof}}}} .
\end{aligned}
\end{equation}
\end{corollary}

\begin{figure*}[!t]
\centerline{\includegraphics[width=440pt]{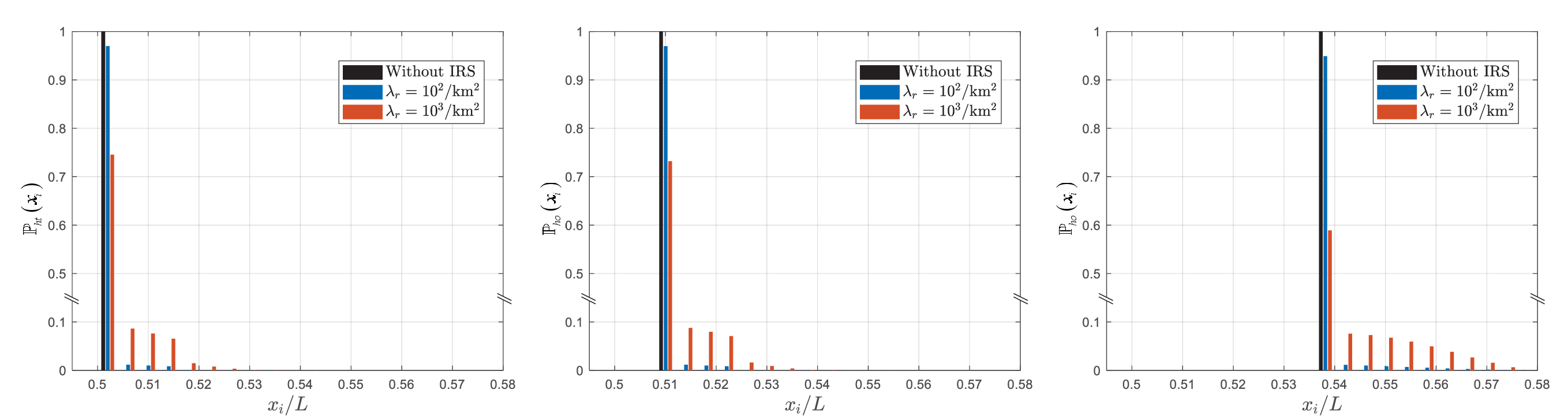}}
\vspace{-0.2cm}
\caption{Probability distributions of HO trigger locations and HO execution locations, where $D=10\rm{m}$: (a) Probability distribution of HO trigger locations; (b) Probability distribution of HO execution locations ($T_t=120\rm{ms}$); (c) Probability distribution of HO execution locations ($T_t=480\rm{ms}$). }
\label{fig1}
\end{figure*}

\begin{figure*}[!t]
\vspace{-0.2cm}
\centerline{\includegraphics[width=440pt]{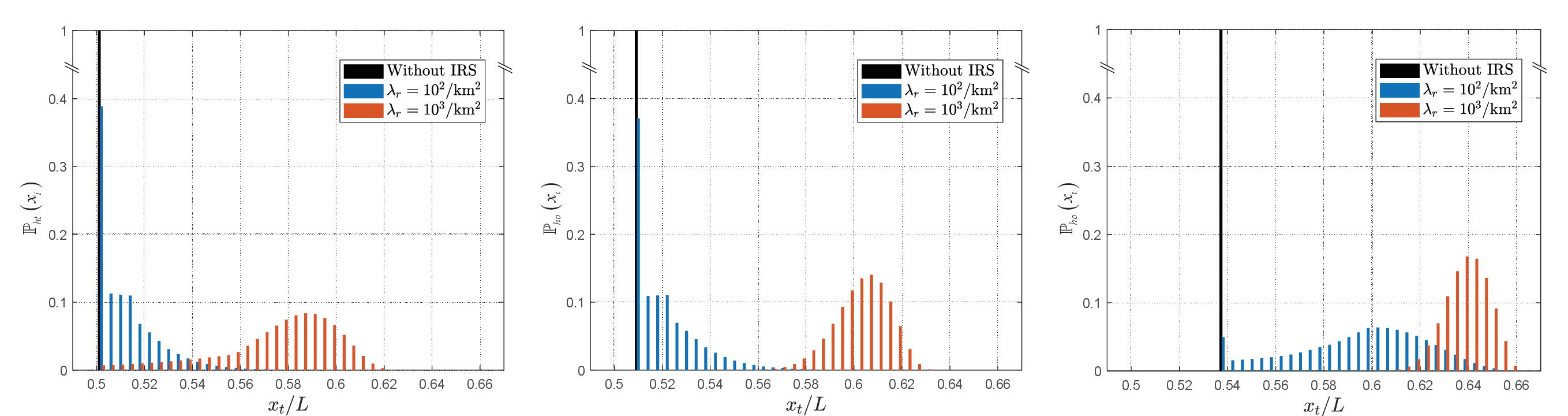}}
\vspace{-0.2cm}
\caption{Probability distributions of HO trigger locations and HO execution locations, where $D=50\rm{m}$: (a) Probability distribution of HO trigger locations; (b) Probability distribution of HO execution locations ($T_t=120\rm{ms}$); (c) Probability distribution of HO execution locations ($T_t=480\rm{ms}$). }
\label{fig1}
\end{figure*}

\vspace{-0.5cm}
\subsection{Ping-Pong Analysis}

According to (4) and (6), a PP event is assumed to occur when the user handovers back to the original BS within the threshold time $T_p$.

As shown in Fig. 3(d), a discrete-time model for the PP event is also proposed, in which the states after the HO is completed are modeled. The PP states are denoted as ${\mathcal{PP}}_0$, ${\mathcal{PP}}_1$, $\cdots$, ${\mathcal{PP}}_{u}$, ${\mathcal{PP}}_t$, and ${\mathcal{PP}}_c$.
The states of the PP are defined as follows. ${\mathcal{PP}}_0$
denotes the state in which the HO is incomplete. ${\mathcal{PP}}_1$ indicates that the HO is complete. If the HO to the original BS is not triggered, the state transitions from ${\mathcal{PP}}_q$ to ${\mathcal{PP}}_{q+1}$ $\left( {q \in \left\{ {1, \cdots ,u - 1} \right\}}, u \!=\! \left\lceil {\frac{{{T_p} - {T_t}}}{{{T_d}}}} \right\rceil \right)$. ${\mathcal{PP}}_{u}$ indicates that the PP is avoided. ${\mathcal{PP}}_{t}$ denotes that the PP condition is just satisfied, and ${\mathcal{PP}}_{c}$ denotes that PP occurs.
Based on the definitions of the PP states, the corresponding
state transition matrix and state vector of the discrete-time model of the PP are constructed as follows:
\begin{lemma}
The state transition matrix of the discrete-time model of the PP in IRS-assisted networks is given by
\begin{equation}
\begin{small}
\begin{aligned}
&{{\rm{\bf{T}}}^{\mathcal{PP}}}\!\!\left( i \right) =\\[-1mm] 
&\left[ {\begin{array}{*{20}{c}}
{\!1 \!-\! p_c^{\mathcal H}\!\left( i \right)}\!&\!{p_c^{\mathcal H}\!\left( i \right)}\!&0& \cdots &0&0&0\\
0&0&\!{1 \!-\! {p^{{\mathcal P}{\mathcal P}}}\!\!\left( i \right)}\!& \cdots &0&\!{{p^{{\mathcal P}{\mathcal P}}}\!\!\left( i \right)}\!&0\\
 \vdots & \vdots & \vdots & \ddots & \vdots & \vdots & \vdots \\
0&0&0&\cdots&\!\!\!{1 \!\!-\!\! {p^{{\mathcal P}{\mathcal P}}}\!\!\left( i \right)}\!\!\!&\! {p^{{\mathcal P}{\mathcal P}}}\!\!\left( i \right)\!&0
\\
0&0&0& \cdots &1&0&0\\
0&0&0& \cdots &0&0&1\\
0&0&0& \cdots &0&0&1
\end{array}} \right]
\end{aligned},
\end{small}
\end{equation}
where its size is $\left( {u + 3} \right) \!\times\! \left( {u + 3} \right)$, the element of the $n$-th row and $m$-th column represents the probability that the $n$-th state transitions to the $m$-th state, the order of states of PP is ${\mathcal{PP}}_0$, ${\mathcal{PP}}_1$, $\cdots$, ${\mathcal{PP}}_{u}$, ${\mathcal{PP}}_t$, ${\mathcal{PP}}_c$, ${p^{\mathcal H}_c}\!\left( i \right)$ indicates the probability of HO completion at the $i$-th measurement moment, and ${p^{{\mathcal {PP}}}}\!\left( i \right)$ denotes the probability of the typical user satisfying the PP condition at $i$-th measurement moment, the expressions of ${p^{\mathcal H}_c}\!\left( i \right)$ and ${p^{\mathcal {PP}}}\!\left( i \right)$ are given by
\begin{equation}
\begin{small}
\begin{aligned}
&p_c^{\mathcal H}\left( i \right) = s_j^{\mathcal H}\left( i \right),
\\
&{p^{{\cal P}{\cal P}}}\!\!\left( i \right) = \!\!\!\sum\limits_{m \in \left\{ {1,2,3,4} \right\}}\!\!\!\! {s_m^{{{\cal I}^t}}\!\!(i){\mathbb P}\left( {{\rm{PP}}\left| {i,{\cal I}_m^t} \right.} \right)} ,
\end{aligned}
\end{small}
\end{equation}
where 
\begin{equation}
\begin{small}
\begin{aligned}
&{\mathbb P}\left( {{\rm{PP}}\left| {i,{\cal I}_m^t} \right.} \right) = {{\mathbb E}_{d,\varphi '}}\left[ {{\mathbb P}\left( {{\rm{PP}}\left| {i,d,\varphi ',{\cal I}_m^t} \right.} \right)} \right],\\
&{\mathbb P}\left( {{\rm{PP}}\left| {i,d,\varphi ',{\cal I}_m^t} \right.} \right) = \left\{ {\begin{array}{ll}
\!\!\!1,&{\eta _{{\rm{PP}},m}}\left( {{x_i},d,\varphi '} \right) > {\gamma _{{\rm{HO}}}}\\
\!\!\! 0,&{\rm{otherwise}}
\end{array}} \right.,\\
&{\eta _{{\rm{PP}},m}}\left( {{x_i},d,\varphi '} \right) \!=\! \left\{ \begin{array}{ll}
\!\!\!{{{\Gamma _{sc}}\left( {{x_o},d,{\varphi _o}} \right)} \mathord{\left/
 {\vphantom {{{\Gamma _{sc}}\left( {{x_o},d,{\varphi _o}} \right)} {{\Gamma _{sc}}\left( {{x_t},d,{\varphi _t}} \right)}}} \right.
 \kern-\nulldelimiterspace} {{\Gamma _{sc}}\left( {{x_t},d,{\varphi _t}} \right)}},&m \in \left\{ {1,3} \right\}\\
\!\!\!{{{\Gamma _{sc}}\left( {{x_o},d,{\varphi _o}} \right)} \mathord{\left/
 {\vphantom {{{\Gamma _{sc}}\left( {{x_o},d,{\varphi _o}} \right)} {{\Gamma _{bf}}\left( {{x_t},d,{\varphi _t}} \right)}}} \right.
 \kern-\nulldelimiterspace} {{\Gamma _{bf}}\left( {{x_t},d,{\varphi _t}} \right)}},&m \in \left\{ {2,4} \right\}
\end{array} \right.\!\!\!.
\end{aligned}
\end{small}
\end{equation}
where pdfs of $d$ and $\varphi '$ are given in \emph{Lemma 3, Lemma 4, and Lemma 5}.
\end{lemma}
\begin{IEEEproof}
According to the definition of the states of the PP, if the HO is executed, the state transitions from $\mathcal{PP}_0$ to $\mathcal{PP}_1$, where the probability of the HO being executed is given by \emph{Theorem 1}. If the PP condition is not met, the state transitions from ${\mathcal{PP}}_1$ to ${\mathcal{PP}}_2$, $\cdots$, ${\mathcal{PP}}_u$, ${\mathcal{PP}}_n$ in turn, otherwise, the state transitions to ${\mathcal{PP}}_t$ and ${\mathcal{PP}}_c$ in turn.

The conditional probabilities of the PP are obtained from (6) and pdfs of the user-IRS distance and angle given in \emph{Lemma 3, Lemma 4, and Lemma 5}. By integrating the conditional probabilities and state vector in \emph{Lemma 2}, ${p^{\mathcal {PP}}}\!\left( i \right)$ is derived.
\end{IEEEproof}
\begin{lemma}
The state vector of the PP in the IRS-assisted networks at the $i$-th measurement is given by
\begin{equation}
\begin{aligned}
&{{\bf{S}}^{\mathcal{PP}}}\!\!\left( i \right) = {{\bf{S}}^{\mathcal{PP}}}\!\!\left( 0 \right) {{\rm{\bf{T}}}^{\mathcal{PP}}}\!\left( 0 \right){{\rm{\bf{T}}}^{\mathcal{PP}}}\!\!\left( 1 \right) \cdots {{\rm{\bf{T}}}^{\mathcal{PP}}}\!\!\left( {i - 1} \right)\\
 &=\! \left[ {s_0^{\mathcal{PP}}\!\left( i \right),s_1^{\mathcal{PP}}\!\left( i \right), \cdots \! ,s_{u}^{\mathcal{PP}}\!\left( i \right),s_t^{\mathcal{PP}}\!\left( i \right),s_c^{\mathcal{PP}}\!\left( i \right)} \right]\!,\\[1mm]
\end{aligned}
\end{equation}
where ${{\bf{S}}^{\mathcal{PP}}}\!\!\left( 0 \right) = \left[ {1,0,0, \cdots ,0} \right]$.
\end{lemma}

Based on the discrete-time model for the PP, the expression of the PP probability is given by the following theorem.
\begin{theorem}
The PP probability of IRS-assisted networks is given by
\begin{equation}
\setlength\abovedisplayskip{3pt}
\setlength\belowdisplayskip{3pt}
\begin{aligned}
{\mathbb{P}}{_{pp}} \!= \!{\Xi _{u + 3}}\!\left[ {{{\bf{S}}^{{\mathcal {PP}}}}\!\!\left( 0 \right){{\rm{\bf{T}}}^{\mathcal{PP}}}\!\!\left( 0 \right){{\rm{\bf{T}}}^{\mathcal{PP}}}\!\!\left( 1 \right) \cdots {{\rm{\bf{T}}}^{\mathcal{PP}}}\!\!\left( {I\! - \!1} \right)} \right] \!=\! s_c^{{\mathcal {PP}}}\!\!\left( I \right),
\end{aligned}
\end{equation}
where ${\Xi_{u+3}}\left[ \cdot  \right]$ indicates the $(u\!+\!3)$-th element of the vector.
\end{theorem}

\begin{corollary}
The average distance from the initial location to the location where the PP event occurs is given by
\begin{equation}
\setlength\abovedisplayskip{3pt}
\setlength\belowdisplayskip{3pt}
\begin{aligned}
{\mathbb{E}}\left[ {{x^{pp}}} \right] \!=\! {{\sum\limits_{i = 1}^I {s_t^{{\mathcal {PP}}}\!\!\left( {{i}} \right) \cdot {x_i}} } \mathord{\left/
 {\vphantom {{\sum\limits_{i = 1}^I {s_t^F\left( {{i}} \right) \cdot {x_i}} } {{P_{hof}}}}} \right.
 \kern-\nulldelimiterspace} {{{\mathbb P}_{pp}}}} .
\end{aligned}
\end{equation}
\end{corollary}

\section{Numerical Results}

In this section, numerical results are provided for the HO, HOF, and PP events. The results are based on the analytical derivations described in the previous sections. Monte Carlo simulations are performed to validate our analysis. Each simulation result is obtained by averaging over $10^4$ randomly generated network topologies in a $10^3$m $\times$ $10^3$m region, with the user executing one HO under each topology.
The following parameters are used if not specific \cite{SingleFad,IRSHO,Hetnet,HetnetFad,b28}: $\lambda_b=10/{\rm{km}}^2$, $\lambda_r=10^3/{\rm{km}}^2$, $P_t=40\rm{dBm}$, $f_c=3\rm{GHz}$, $c = 3\times 10^8 \rm{m/s}$, $\alpha =4$, $D=50\rm{m}$, $N=100$, $v=20\rm{m/s}$, $T_d=10\rm{ms}$, ${\gamma _{{\rm{HO}}}}=0\rm{dB}$, $T_t=480\rm{ms}$, $T_p=1\rm{s}$, $Q_{out}=-8\rm{dB}$, ${r_o}\!\! =\! -r_t\!=\!\! \int_0^\infty \!\!\! {\int_0^{2\pi }\!\!\! {\sin \phi {r^2}{\lambda _b}{e^{ - {\lambda _b}\pi {r^2}}}\!\!{\rm{d}}\phi {\rm{d}}r} }$.

\subsection{Results of Handover}

In this section, we focus on the effect of IRS on the HO trigger and execution locations. In particular, we quantitatively investigate the hysteresis effect of IRS-related parameters on HO using numerical results.

Figs. 4 and 5 show the probability distributions of the HO trigger and execution locations under different IRS serving distances $D$ and IRS densities $\lambda_r$. Compared to the network without IRS, the HO trigger locations and HO execution locations of IRS-assisted networks become irregular, as evidenced by the probabilities that triggering HO and executing HO being present at multiple locations. The HO trigger locations shift toward the target BS as the IRS density $\lambda_r$ and IRS serving distance $D$ increase. Specifically, the probability of HO triggering is the highest at ${{x_i}\mathord{\left/ \right.
\kern-\nulldelimiterspace} L}\!=\! 0.586$ when $D = 50\rm{m}$, $\lambda_r=10^3/\rm{km^2}$, and the probability of HO triggering is higher than 90\% when ${{x_i}\mathord{\left/ \right.
\kern-\nulldelimiterspace} L}$ is in the interval from 0.55 to 0.61. Meanwhile, approximately 97\% of the HOs are triggered at ${{x_i}\mathord{\left/ \right.
\kern-\nulldelimiterspace} L}\!=\! 0.503$ when $D = 10\rm{m}$, $\lambda_r=10^2/\rm{km^2}$. For the HO execution location, increasing the TTT duration delays the HO execution, which facilitates the avoidance of PP but also presents the risk of HOF as the user needs to receive the HO command from the original BS. Specifically, the IRS further delays HO locations, e.g., a significant number of HO executions are distributed around ${{x_i}\mathord{\left/ \right.
\kern-\nulldelimiterspace} L}\!=\! 0.606$, although the HO has an 80\% probability of triggering at ${{x_i}\mathord{\left/ \right.
\kern-\nulldelimiterspace} L} \!<\! 0.52$, when $D = 50\rm{m}$, $\lambda_r=10^2/\rm{km^2}$.

Fig. 6 shows the effect of the number of IRS elements on the HO trigger location for different IRS serving distances, IRS densities, and HO margins. When ${\gamma _{{\rm{HO}}}}\!<\!1$, the HO trigger locations are shifted toward the original BS; otherwise, they are shifted toward the target BS. The IRS has a more significant effect on the HO trigger locations when ${\gamma _{{\rm{HO}}}}<1$, e.g., when $N=100$, $D=50\rm{m}$, $\lambda_r=10^3/\rm{km^2}$, ${\mathbb{E}{\left[ {{x^{ht}}} \right]} \mathord{\left/ \right.
\kern-\nulldelimiterspace} L}$ rises by 26.9\% for the case of ${\gamma _{{\rm{HO}}}}=-2\rm{dB}$ compared to the case without IRS, while ${\mathbb{E}{\left[ {{x^{ht}}} \right]} \mathord{\left/ \right.
\kern-\nulldelimiterspace} L}$ only rises by  8.2\% as ${\gamma _{{\rm{HO}}}}=2\rm{dB}$. Even when ${\gamma _{{\rm{HO}}}}=2\rm{dB}$, $\lambda_r=10^2/\rm{km^2}$, the HO location does not change significantly with IRS. Moreover, the HO trigger location will not continue to significantly change with increasing $N$ when $N$ reaches a certain value. This value increases as $D$ and $\lambda_r$ increases and $\gamma_{\rm{HO}}$ decreases, e.g., for the case of $\lambda_r={10^3}/\rm{km^2}$ and $\gamma_{\rm{HO}}=2\rm{dB}$, ${\mathbb{E}{\left[ {{x^{ht}}} \right]} \mathord{\left/ \right.
\kern-\nulldelimiterspace} L}$ does not change significantly with $N$ when $N$ reaches 7 and 300 for $D=20\rm{m}$ and $D=50\rm{m}$, respectively.

Fig. 7 shows the impact of the IRS serving distances on the HO trigger location under different IRS densities, numbers of IRS elements, and HO margins. Different IRS densities do not result in different HO trigger locations when $D$ is low. A sufficient value of $D$ is required to bring about differences between different IRS densities, whereas the effect of $N$ is not limited by $D$. For the case of ${\gamma _{{\rm{HO}}}}\!=\!2\rm{dB}$, IRS only affects the HO trigger location when $D$ is greater than 15m. Similar to the trend of the HO trigger locations with $N$, the delay in HO triggering slows down when $D$ increases. The change in the HO trigger location requires extra attention in the parameter intervals where it varies sharply.

\begin{figure}[!t]
\vspace{-0.2cm}
\centerline{\includegraphics[width=0.4\textwidth]{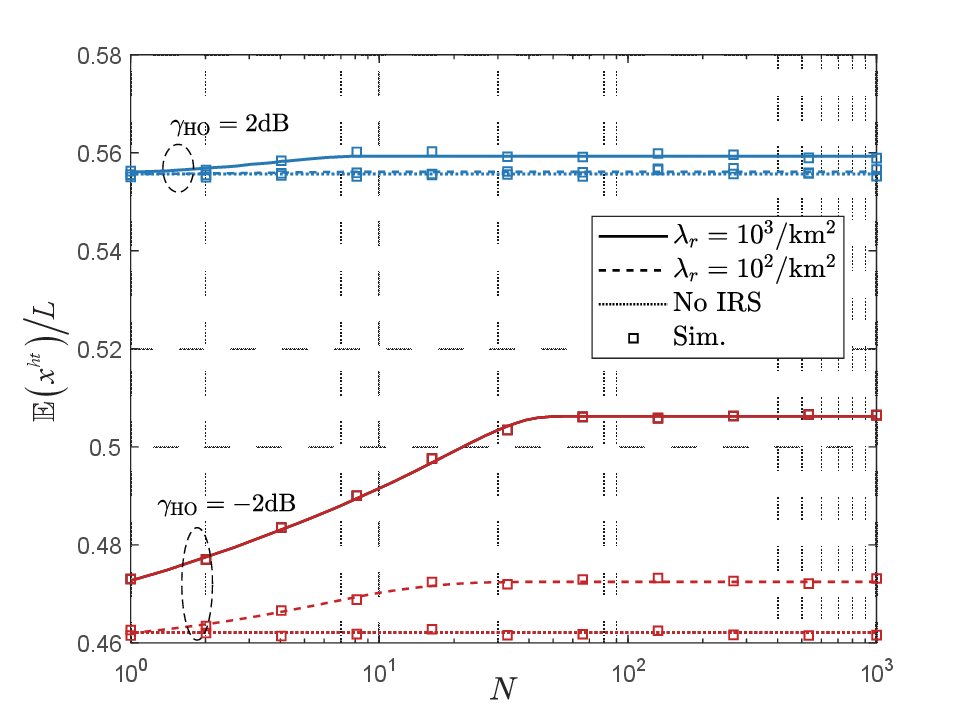}}
\vspace{-0.1cm}
\centerline{\includegraphics[width=0.4\textwidth]{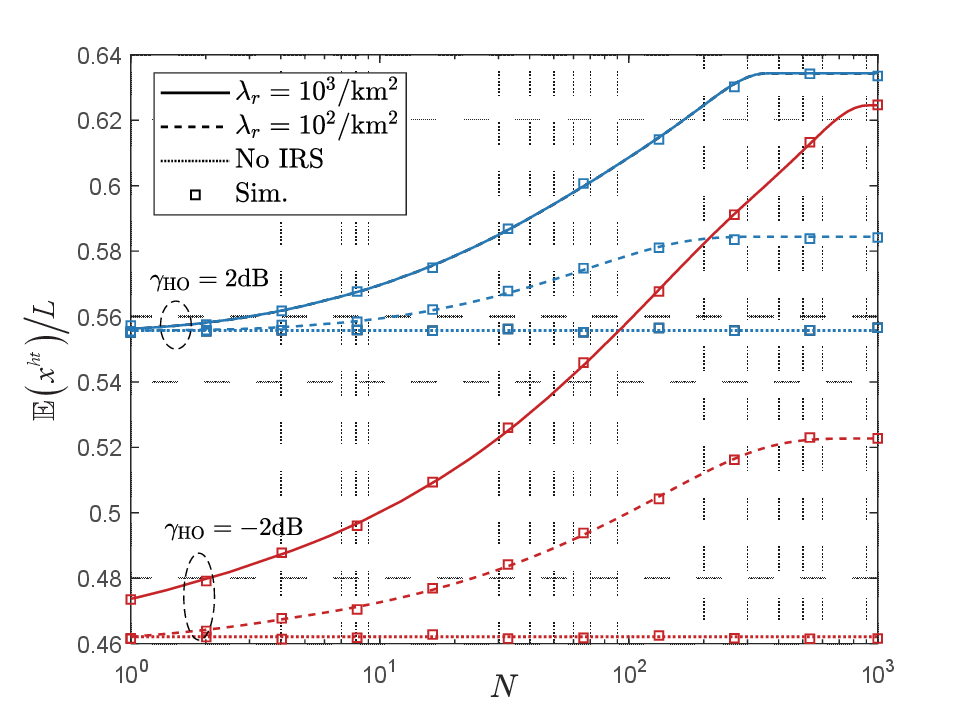}}
\vspace{-0.3cm}
\caption{Ratio of the average distance from the initial location to the location where the HO is triggered to $L$ as a function of the number of IRS elements $N$ under different HO margins ${\gamma _{{\rm{HO}}}}$: (a) $D= 20 \rm{m}$; (b) $D= 50 \rm{m}$.}
\label{fig1}
\vspace{-0.2cm}
\end{figure}

\begin{figure}[!t]
\vspace{-0.2cm}
\centerline{\includegraphics[width=0.4\textwidth]{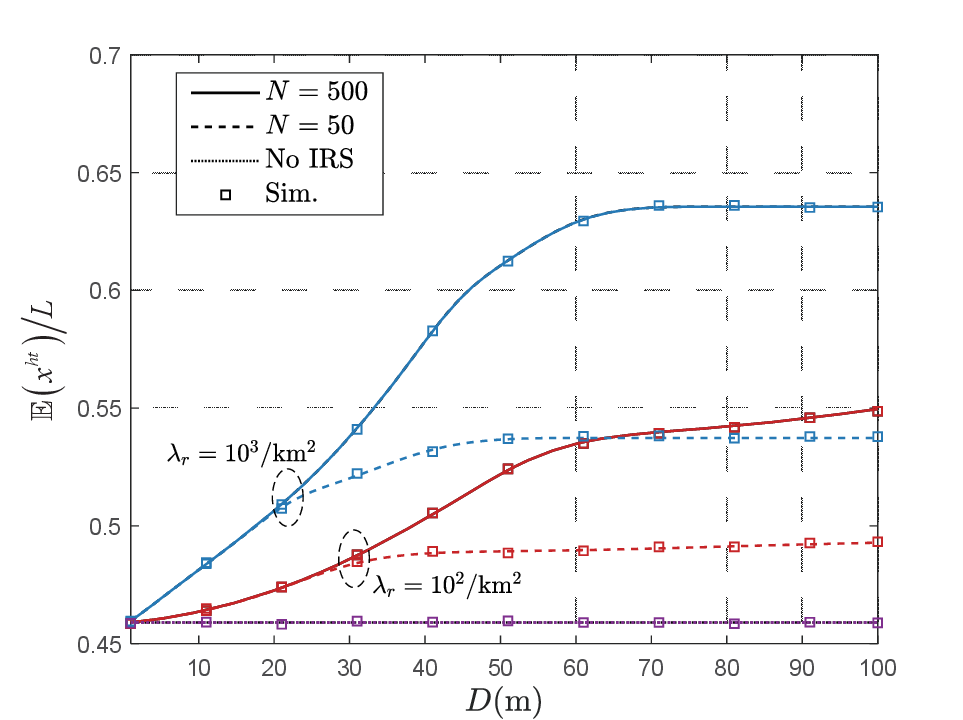}}
\vspace{-0.1cm}
\centerline{\includegraphics[width=0.4\textwidth]{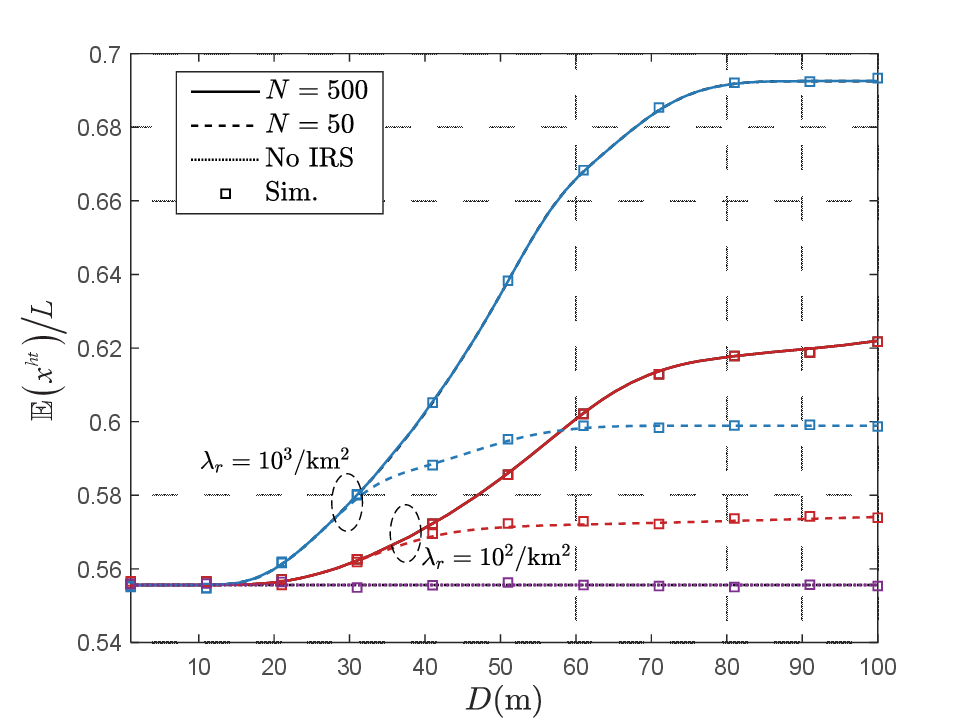}}
\vspace{-0.3cm}
\caption{Ratio of the average distance from the initial location to the location where the HO is triggered to $L$ as a function of the IRS serving distance $D$ under different HO margins ${\gamma _{{\rm{HO}}}}$: (a) ${\gamma _{{\rm{HO}}}}=-2\rm{dB}$; (b) ${\gamma _{{\rm{HO}}}}=2\rm{dB}$.}
\label{fig1}
\vspace{-0.2cm}
\end{figure}

\subsection{Results of Handover Failure and Ping-Pong}
In this section, we focus on analyzing the probabilities of HOF and PP in IRS-assisted networks based on the setup mentioned above. Specifically, we discuss the impact of IRS configuration parameters on HOF and PP events. We let ${\gamma _{{\rm{HO}}}}$ be $-2\rm{dB}$ to observe the trends in the PP probability.

Fig. 8 shows the probabilities of HOF and PP as functions of the IRS serving distance $D$ for different IRS densities and numbers of IRS elements. For the HOF, there exists a $D^*$ that makes the HOF most severe, and $D^*$ takes different values for different IRS element numbers and IRS densities. This is owing to the delay in the HO trigger location after the increase in $D$ (as shown in Fig. 6) and the increase in interference to the user from the target BS after TTT. However, a sufficiently large IRS serving distance ensures signal strength from the original BS and helps avoid the SIR from falling below $Q_{out}$. For the PP event, since the user can connect to the IRS in the new cell after the HO, it helps the user stay in the new cell. When the IRS density is large ($\lambda_r=10^3/\rm{km^2}$), the probability that a user can connect to an IRS in the new cell rises rapidly with $D$ and the user is close to the IRS. Thus, the PP probability decreases rapidly with $D$ and a large $N$ is not required to avoid PP in this case.

Fig. 9 shows the probabilities of HOF and PP as functions of IRS density for different numbers of IRS elements. There are trade-offs between HOF and PP in IRS density and IRS element number, where a higher IRS density or a larger IRS element number indicates more HOFs but fewer PPs. Therefore, when the IRS density and number of IRS elements are large, HOF should be considered first; otherwise, PP should be avoided. In particular, the PP probability decreases by 46\%, and the HOF probability increases by 91\% when $N$ reaches 100 and $\lambda_r=500/\rm{km^2}$, but with only a 7\% increase in HOF and a 7\% drop in PP as $N$ continues to rise to $800$. As the IRS density increases, the trend of HOF increase slows down, yet the trend of the PP decrease remains nearly constant until its probability is less than 0.1.

\begin{figure}[!t]
\vspace{-0.2cm}
\centerline{\includegraphics[width=0.4\textwidth]{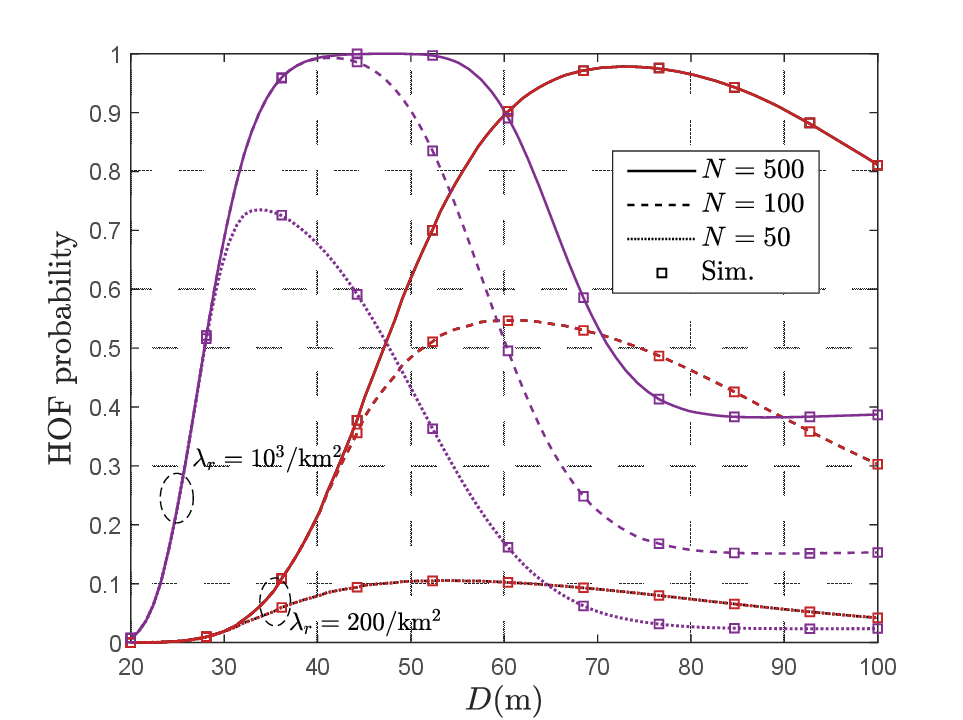}}
\centerline{\includegraphics[width=0.4\textwidth]{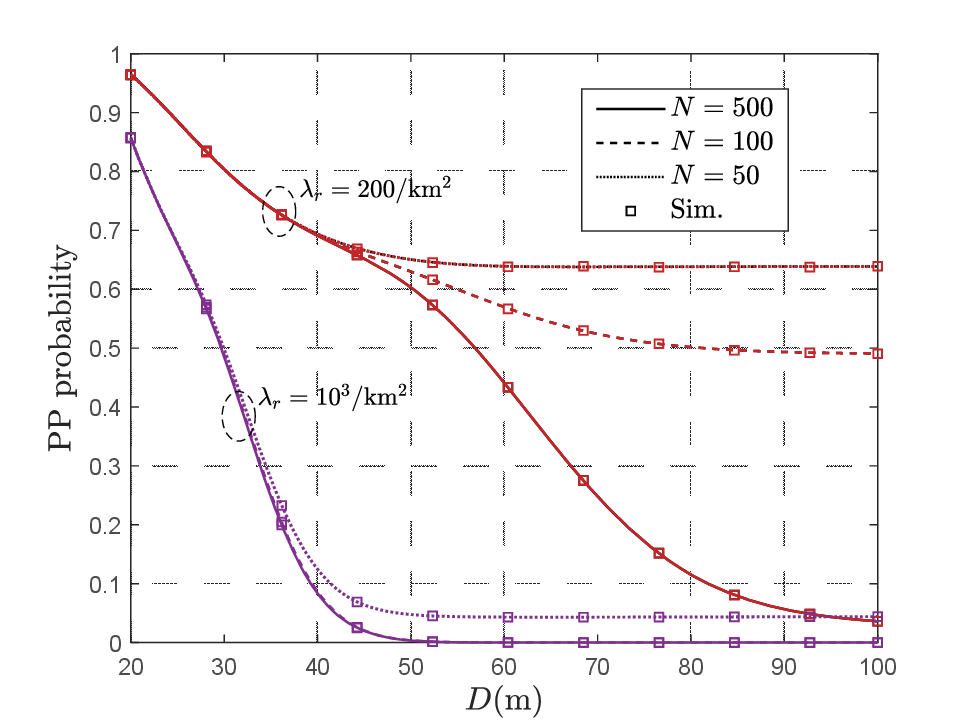}}
\vspace{-0.3cm}
\caption{HOF probability and PP probability as a function of the IRS serving distance $D$ under different IRS densities $\lambda_r$ and numbers of IRS elements $N$: (a) HOF probability; (b) PP probability.}
\label{fig1}
\vspace{-0.2cm}
\end{figure}

\subsection{Optimal Settings of TTT and HO Margin}

The HO parameters (i.e., TTT and HO margin) are easier to configure than the network deployment parameters (e.g., IRS density), and changes in the HO parameters do not significantly affect the performance of other networks (e.g., spectral efficiency). Therefore, selecting the optimal HO parameters is a convenient and effective method to improve the HO performance.
In this section, the optimal HO parameters in IRS-assisted networks are mined to minimize the probabilities of HOF and PP. The trends in the optimal HO parameters under different IRS configurations are also discussed.

Fig. 10 shows the trade-off between the HOF and PP when the HO parameters are set under different IRS element numbers. Greater TTT and HO margins postpone HO execution, avoiding some PPs but leading to HOFs. Thus, a relatively high TTT requires a low HO margin to balance HOF and PP. As $N$ increases, the HO locations shift toward the target BS. Therefore, the appropriate HO margin under different TTTs decreases with increasing $N$ to back off the HO locations. Moreover, the optimal HO parameters decrease as $N$ increases. In particular, with a target of less than 5\% for both HOF and PP, when TTT is configured as 240ms, the ranges of suitable HO margins are $[-2.2,1.5]\rm{dB}$ and $[-6,-3]\rm{dB}$ for $N=20$ and $N=200$, respectively. However, when $N=500$, only $\gamma_{\rm{HO}}\!=\! -6.2\rm{dB}$ satisfies this requirement; other TTT configurations must be considered in this case.

Fig. 11 shows the trade-off between HOF and PP when setting the HO parameters under different IRS serving distances. Combining the conclusions of Figs. 6 and 7, increasing $D$ delays the HO location, but a sufficient $D$ prevents HOF. Therefore, as $D$ increases, the optimal HO margin interval first shifts to smaller values and then expands. In particular, with a target of less than 5\% for both HOF and PP, when TTT is configured as 240 ms, the reasonable HO margin intervals are $[-0.2,3]\rm{dB}$, $[-5.7,-2.4]\rm{dB}$, and $[-6,0.8]\rm{dB}$ for cases of $D=10\rm{m}$, $D=50\rm{m}$, and $D=100\rm{m}$, respectively.

A heatmap of the HOF and PP is depicted in Fig. 12 for detailed parameter settings under different IRS densities. The trends of HOF and PP with TTT and the HO margins are opposite. As the IRS density increases, the appropriate HO margin tends to decrease. For instance, when both the HOF and PP probabilities are desired to be less than 0.1\%, the broadest TTT choices occur at $\gamma \!=\! 0\rm{dB}$ and $\gamma \!=\! -6\rm{dB}$ for the cases of $\lambda_r\!=\!200/\rm{km^2}$ and $\lambda_r\!=\!10^3/\rm{km^2}$, respectively. The corresponding parameters are $\left\{ {{T_t} = 100{\rm{ms}},{\gamma _{{\rm{HO}}}} = 0{\rm{dB}}} \right\}$ and $\left\{ {{T_t} = 200{\rm{ms}},{\gamma _{{\rm{HO}}}} = 0{\rm{dB}}} \right\}$ for the case of $\lambda_r\!=\!200/\rm{km^2}$, $\left\{ {{T_t} = 300{\rm{ms}},{\gamma _{{\rm{HO}}}} = - 6{\rm{dB}}} \right\}$, $\left\{ {{T_t} = 400{\rm{ms}},{\gamma _{{\rm{HO}}}} = - 6{\rm{dB}}} \right\}$, and $\left\{ {{T_t} = 500{\rm{ms}},{\gamma _{{\rm{HO}}}} = - 6{\rm{dB}}} \right\}$ for the case of $\lambda_r\!=\!10^3/\rm{km^2}$.

\begin{figure}[!t]
\vspace{-0.2cm}
\centerline{\includegraphics[width=0.4\textwidth]{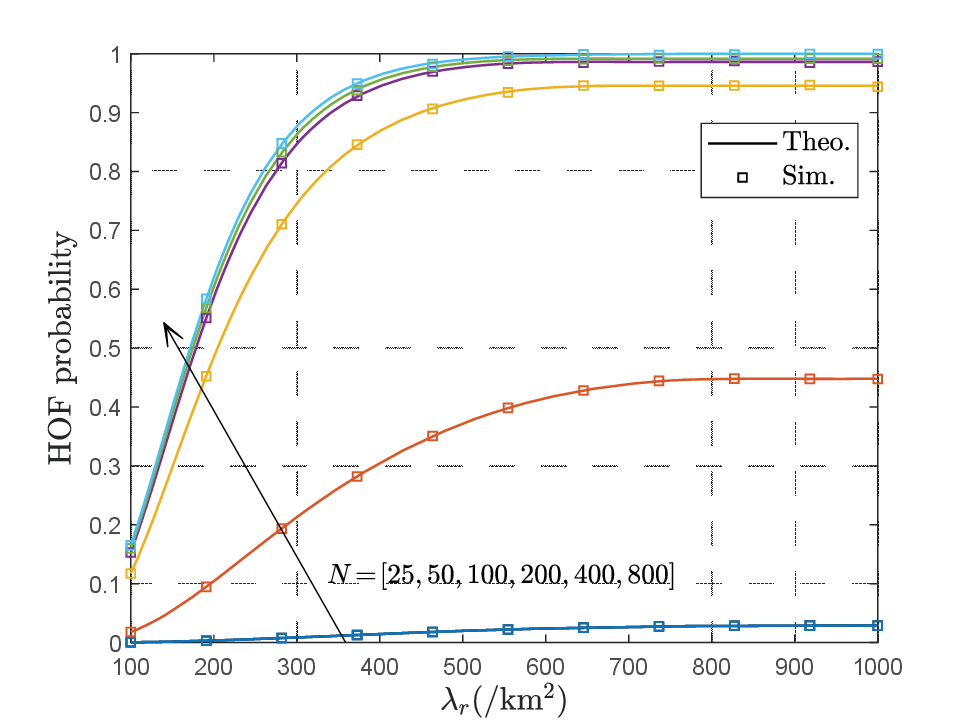}}
\centerline{\includegraphics[width=0.4\textwidth]{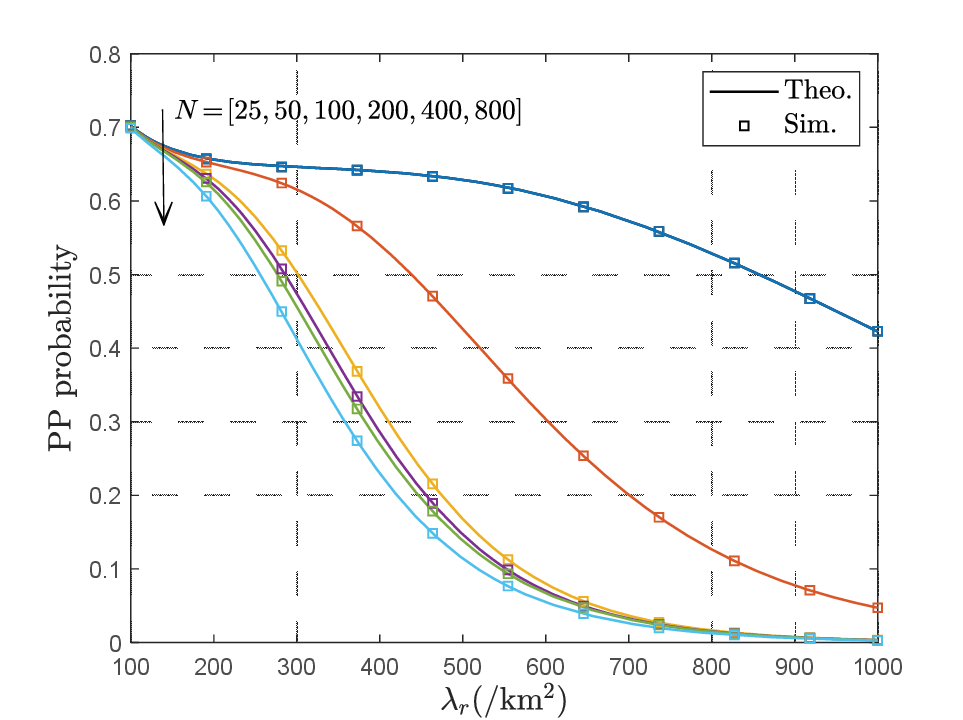}}
\vspace{-0.3cm}
\caption{HOF probability and PP probability as a function of the IRS density $\lambda_r$ under different numbers of IRS elements $N$: (a) HOF probability; (b) PP probability.}
\label{fig1}
\vspace{-0.2cm}
\end{figure}

\begin{figure*}[!t]
\vspace{-0.2cm}
\centerline{\includegraphics[width=440pt]{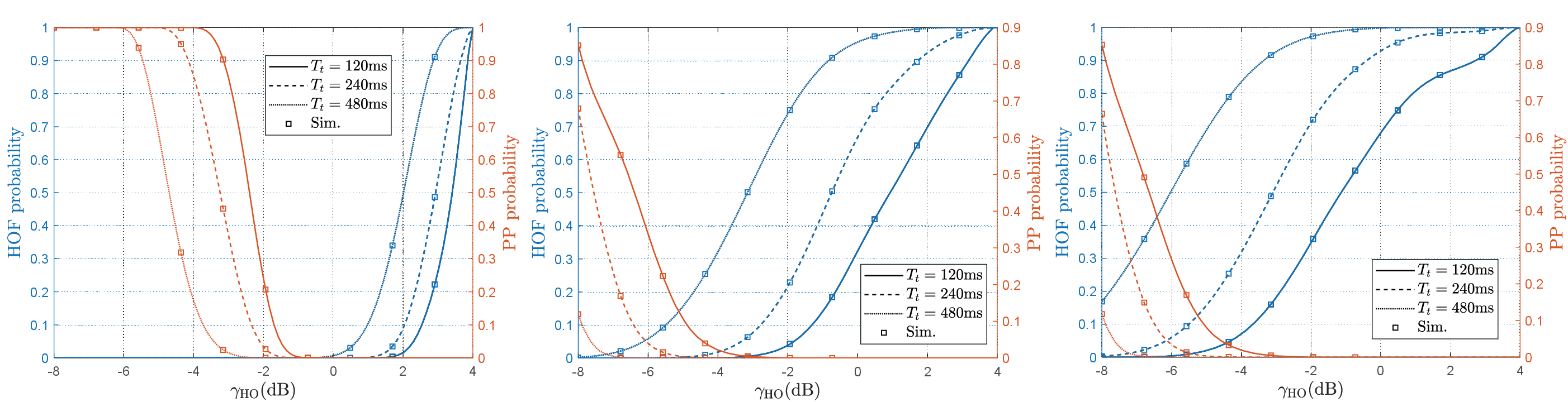}}
\vspace{-0.2cm}
\caption{Relationship between the handover parameters and handover failure probability, ping-pong probability, where the blue curves represent handover failure probabilities and the red curves represent ping-pong probabilities: (a) $N$=20; (b) $N$=200; (c) $N$=500.}
\vspace{-0.2cm}
\label{fig1}
\end{figure*}
\begin{figure*}[!t]
\vspace{-0.2cm}
\centerline{\includegraphics[width=440pt]{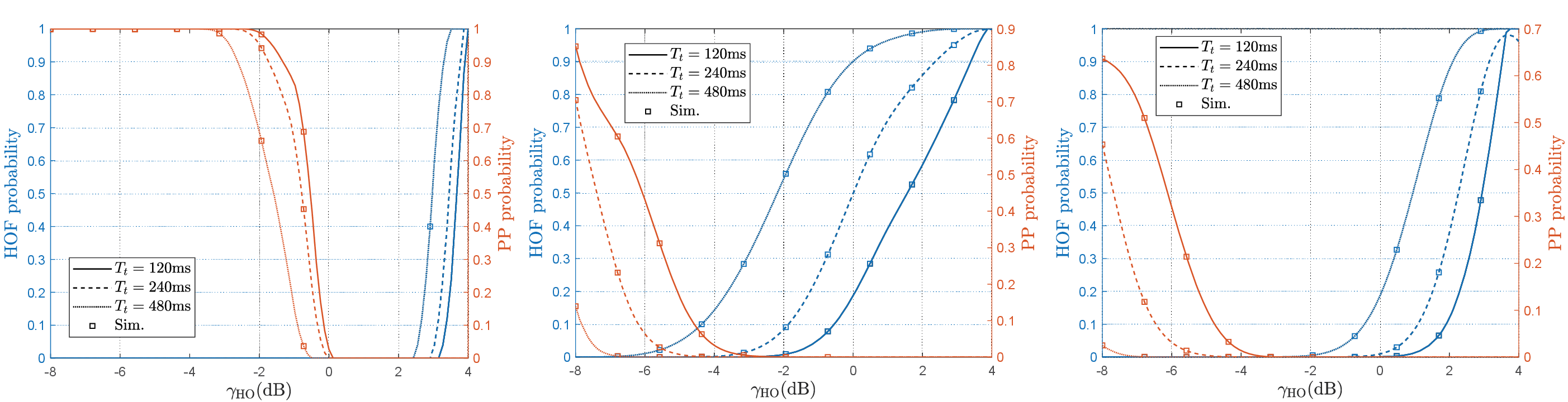}}
\vspace{-0.2cm}
\caption{Relationship between the handover parameters and handover failure probability, ping-pong probability, where the blue curves represent handover failure probabilities and the red curves represent ping-pong probabilities: (a) $D$=10m; (b) $D$=50m; (c) $D$=100m.}
\vspace{-0.2cm}
\label{fig1}
\end{figure*}

\begin{figure}[!t]
\vspace{-0.2cm}
\centerline{\includegraphics[width=0.42\textwidth]{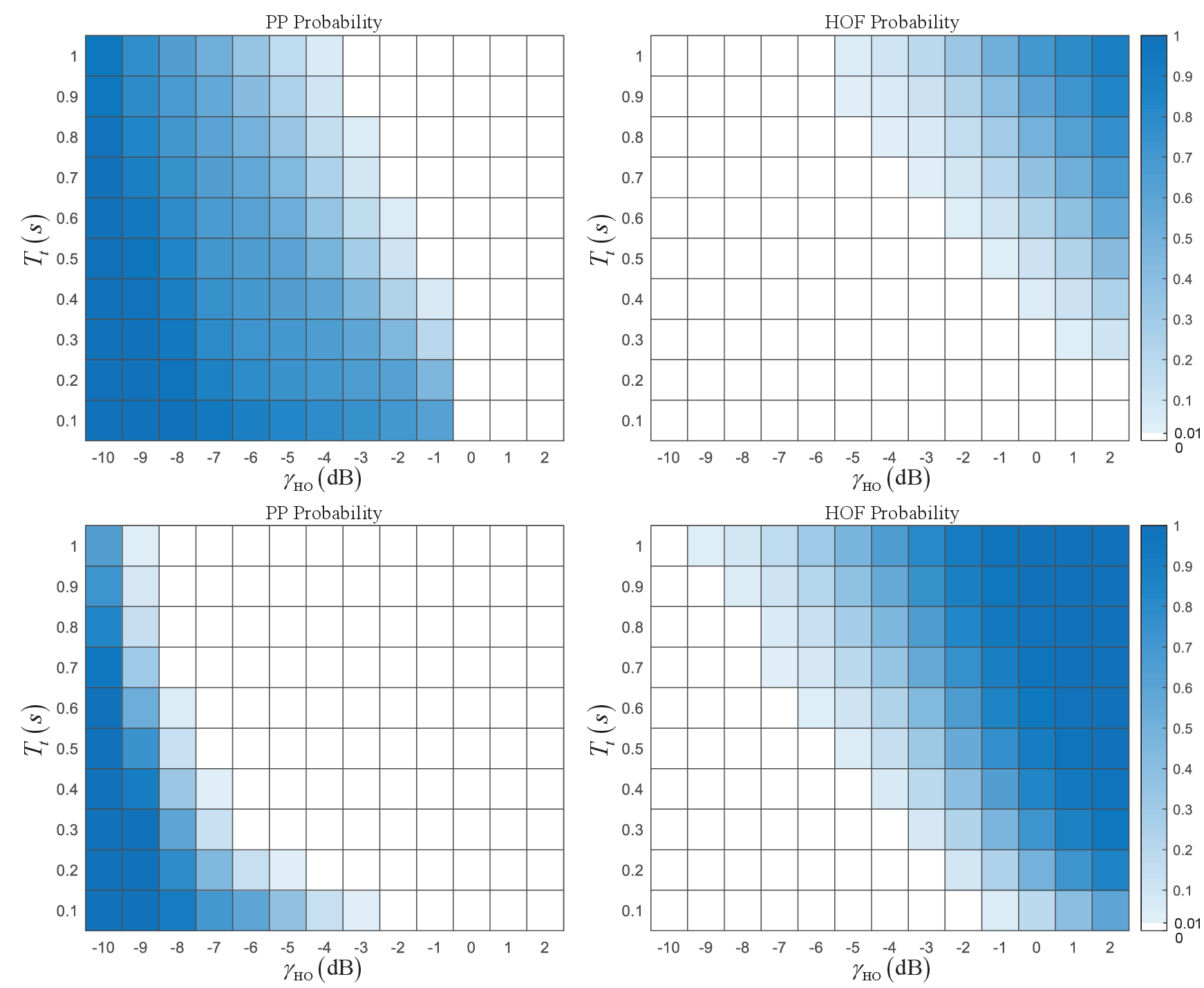}}
\vspace{-0.2cm}
\caption{Heatmap of the handover failure probability and ping-pong probability under different handover parameters: (a) PP probability for $\lambda_r = 200/\rm{km^2}$; (b) HOF probability for $\lambda_r = 200/\rm{km^2}$; (c) PP probability for $\lambda_r = 10^3/\rm{km^2}$; (d) HOF probability for $\lambda_r = 10^3/\rm{km^2}$.}
\vspace{-0.6cm}
\label{fig1}
\end{figure}

\vspace{-0.2cm}
\section{Conclusion}

To analyze the effect of IRS on the HO process and minimize HOFs and PPs in IRS-assisted networks, the HO process in IRS-assisted networks was theoretically modeled via a discrete-time model, where the exact HO location with IRS signal enhancement was captured, and HO states under IRS channel fluctuations were explicitly tracked. The probability distributions of HO triggering and HO execution over user trajectories, and the probabilities of HOF and PP were deduced. The analytical expressions were highly accurate when compared to the simulation results. Furthermore, we provide a comprehensive study of the HO performance trend using the IRS configuration parameters and determine the optimal HO parameters (i.e., TTT and HO margin) in IRS-assisted networks. This study could be extended in several methods. For instance, optimization algorithms for HO parameters can be designed based on analysis, or more practical factors, such as blockage effects, can be introduced and analyzed.

\vspace{-0.2cm}
\begin{appendix}
\vspace{-0.2cm}
\subsection{Proof of Lemma 1}
\vspace{-0.1cm}
As shown in Fig. 13, $l$ is the straight line equidistant from the original BS and the target BS, $l$ intersects the user trajectory at point $\mathbf{x}_{\rm{mid}}$. The two regions divided by $l$ are denoted $\mathcal{B}^o$ for the part close to the original BS and $\mathcal{B}^t$ for the part close to the target BS, the regions centered on the user's location at the $(i-1)$-th and $i$-th measurement moments ($\mathbf{x}_{i-1}$ and $\mathbf{x}_{i}$), with $D$ as the radius, are denoted as $\mathcal{A}_{i-1}$ and $\mathcal{A}_{i}$, respectively.

Consider the IRS connection states of the original BS (i.e., $k\!=\!o$) for example.
When the user has not handed over the target BS, the user can connect to the IRS in $\mathcal{B}_o$.
To consider the correlation of adjacent moments, we focus on the following three regions when deriving the state transition probabilities: (i) the region where the IRS may exist at the $i$-th measurement moment, which is denoted as ${{\mathcal A}'_i}\!=\!{{\mathcal A}_i}\!\cap\! {{\mathcal B}^o}$ and (ii) the region where the IRS may exist at $(i\!-\!1)$-th measurement moment, denoted by ${{\mathcal A}'_{i-1}}\!=\!{{\mathcal A}_{i-1}}\!\cap\! {{\mathcal B}^o}$; and (iii) the overlap between ${{\mathcal A}'_i}$ and ${{\mathcal A}'_{i-1}}$, which is denoted as ${\mathcal A}_{i,i - 1}^ \cap \!=\!{{\mathcal A}'_{i-1}}\!\cap\!{{\mathcal A}'_{i}}$.
Thus, we obtain
\begin{equation}
\setlength\abovedisplayskip{3pt}
\setlength\belowdisplayskip{3pt}
\begin{small}
\begin{aligned}
&p_{1,1}^{{{\mathcal I}^o}}\left( i \right) = p_{3,1}^{{{\mathcal I}^o}}\left( i \right)\\
 &= {\mathbb P}\left( {N\left| {{{{\mathcal A}'}_i}} \right| = 0\left| {N\left| {{{{\mathcal A}'}_{i - 1}}} \right| = 0} \right.} \right) = {\mathbb P}\left( {N\left| {{{{\mathcal A}'}_i}\backslash {\mathcal A}_{i,i - 1}^ \cap } \right| = 0} \right),
\\[0mm]
&p_{2,3}^{{{\mathcal I}^o}}\left( i \right) = p_{4,3}^{{{\mathcal I}^o}}\left( i \right) \\ &={\mathbb P}\left( {N\left| {{\mathcal A}_{i,i - 1}^ \cap } \right| = 0\left| {N\left| {{{\mathcal A'}_{i - 1}}} \right| \ne 0} \right.} \right) {\mathbb P}\left( {N\left| {{{\mathcal A'}_i}\backslash {\mathcal A}_{i,i - 1}^ \cap } \right| = 0} \right)\\
& = \left[ {1 \!-\! \frac{{{\mathbb P}\!\left( {N\left| {{\mathcal A}_{i,i - 1}^ \cap } \right| \ne 0,N\left| {{{\mathcal A'}_{i - 1}}} \right| \ne 0} \right)}}{{{\mathbb P}\left( {N\left| {{{\mathcal A'}_{i - 1}}} \right| \ne 0} \right)}}} \right]  {\mathbb P}\!\left( {N\left| {{{\mathcal A'}_i}\backslash {\mathcal A}_{i,i - 1}^ \cap } \right| \!=\! 0} \right)\\
&\mathop  = \limits^{\left( a \right)} \left[ {1 \!-\! \frac{{{\mathbb P}\left( {N\left| {{\mathcal A}_{i,i - 1}^ \cap } \right| \ne 0} \right)}}{{{\mathbb P}\left( {N\left| {{{\mathcal A'}_{i - 1}}} \right| \ne 0} \right)}}} \right] {\mathbb P}\left( {N\left| {{{\mathcal A'}_i}\backslash {\mathcal A}_{i,i - 1}^ \cap } \right| = 0} \right),\\
&p_{1,2}^{{{\mathcal I}^o}}\!\!\left( i \right) \!=\! p_{3,2}^{{{\mathcal I}^o}}\!\!\left( i \right)\! =\! 1 \!-\! p_{1,1}^{{{\mathcal I}^o}}\!\!\left( i \right),p_{2,3}^{{{\mathcal I}^o}}\!\!\left( i \right) \!=\! p_{4,3}^{{{\mathcal I}^o}}\!\!\left( i \right) \!=\! 1 \!-\! p_{2,4}^{{{\mathcal I}^o}}\!\!\left( i \right)\!,
\end{aligned}
\end{small}
\end{equation}
where $N\left|  \cdot  \right|$ is the counting measure and (a) follows ${\mathcal A}_{i,i - 1}^ \cap  \subseteq {{{\mathcal A}'}_{i - 1}}$. According to the null probability of a 2-D PPP, the probability that there is no IRS in region $\mathcal{A}$ is  
\begin{equation}
\setlength\abovedisplayskip{3pt}
\setlength\belowdisplayskip{3pt}
\begin{small}
{\mathbb P}\left( {N\left| {\mathcal A} \right| = 0} \right) = {e^{ - {\lambda _r}\left| {\mathcal A} \right|}},
\end{small}
\end{equation}
where $\left|  \cdot  \right|$ is Lebesgue measure for sets. Then, we focus on calculating the area of regions of interest. Since ${\mathcal A}_{i,i - 1}^ \cap  \subseteq {{{\mathcal A}'}_{i}}$, we have $\left| {{{\mathcal A'}_i}\backslash {\mathcal A}_{i,i - 1}^ \cap } \right| = \left| {{\mathcal A'}_i} \right| - \left| {\mathcal A}_{i,i - 1}^ \cap \right|$, where $\left| {{\mathcal A'}_i} \right|$ and $\left| {\mathcal A}_{i,i - 1}^ \cap \right|$ are given by
\begin{equation}
\setlength\abovedisplayskip{3pt}
\setlength\belowdisplayskip{3pt}
\begin{small}
\begin{aligned}
&\left| {{{{\cal A}'}_i}} \right| = \left\{ \begin{array}{ll}
\left| {{{\cal A}_i}} \right|,&0 \le {x_i} \le {x_{{\rm{mid}}}} - \frac{D}{{\sin \theta }}\\
\left| {{{\cal A}_i}} \right| - \left| {{{\cal A}_i} \cap {{\cal B}^t}} \right|,&{x_{{\rm{mid}}}} - \frac{D}{{\sin \theta }} < {x_i} \le {x_{{\rm{mid}}}} + \frac{D}{{\sin \theta }}\\[0.5mm]
0,&{x_{{\rm{mid}}}} + \frac{D}{{\sin \theta }} < {x_i}
\end{array} \right.,\\
&\left| {{\mathcal A}_{i,i - 1}^ \cap } \right| = \left\{ \begin{array}{ll}
\left| {{{\mathcal A}_{i - 1}} \cap {{\mathcal A}_i}} \right|,&\!\!\!\!\!\!\!0 \le {x_i} \le {x_{{\rm{mid}}}} - \frac{D}{{\sin \theta }} + \Delta x\\
\left| {\left( {{{\mathcal A}_{i - 1}} \cap {{\mathcal A}_i}} \right) \backslash {{\mathcal B}^t}} \right|,\\&\!\!\!\!\!\!\!\!\!\!\!\!\!\!\!\!\!\!\!\!\!\!\!\!\!\!\!\!\!\!\!{x_{{\rm{mid}}}} - \frac{D}{{\sin \theta }} + \Delta x < {x_i} \le {x_{{\rm{mid}}}} + \frac{D}{{\sin \theta }}\\[1mm]
0,&\ \ \ \ \ \ \ \ \ \ {x_{{\rm{mid}}}} + \frac{D}{{\sin \theta }} < {x_i}
\end{array} \right.,
\end{aligned}
\end{small}
\end{equation}
where $\left| {\left( {{{\mathcal A}_{i - 1}}\! \cap \!{{\mathcal A}_i}} \right)\! \backslash {{\mathcal B}^t}} \right|$ is calculated by integration because deriving its closed-form expression is not tractable, and a polar coordinate is established with the user location at the $i$-th measurement moment as the origin and the conditions for $\left( {\rho ,\theta } \right) \in {\left( {{{\mathcal A}_{i - 1}} \cap {{\mathcal A}_i}} \right) \backslash {{\mathcal B}^t}}$ are: (i) $\left( {\rho ,\theta } \right) \in {{\mathcal A}_i}$, i.e., $0 < \rho  < D$, (ii) $\left( {\rho ,\theta } \right) \in {{\mathcal A}_{i-1}}$, i.e., ${0 < \sqrt {\Delta {x^2} + {\rho ^2} + 2\rho \Delta x\cos \varphi }  < D}$, (iii) $\left( {\rho ,\theta } \right) \in {\mathcal B}^o$, i.e., ${\left[ {\rho \sin \varphi  + \rho \tan \theta \cos \varphi  - \tan \theta \left( {{x_{{\rm{mid}}}} - {x_i}} \right)} \right] \left( {\theta  - \frac{\pi }{2}} \right) > 0}$. By calculating the area of regions and substituting (41), (42) into (40), (9) is derived, where $\left| {{\mathcal A'}_i} \right|$ and $\left| {\mathcal A}_{i,i - 1}^ \cap \right|$ are denoted as $S_i^o$ and $S_i^{o,\cap}$ for the sake of simplicity.

Using the same procedures, the state transition probabilities of the IRS connection states of the target BS (i.e., $k=t$) can be obtained, where $\mathcal{A}'_{i}$ and $\mathcal{A}'_{i-1}$ need to be replaced by $\mathcal{A}_{i} \cap \mathcal{B}^t$ and $\mathcal{A}_{i-1} \cap \mathcal{B}^t$.

\begin{figure}[!t]
\centerline{\includegraphics[width=0.65 \linewidth]{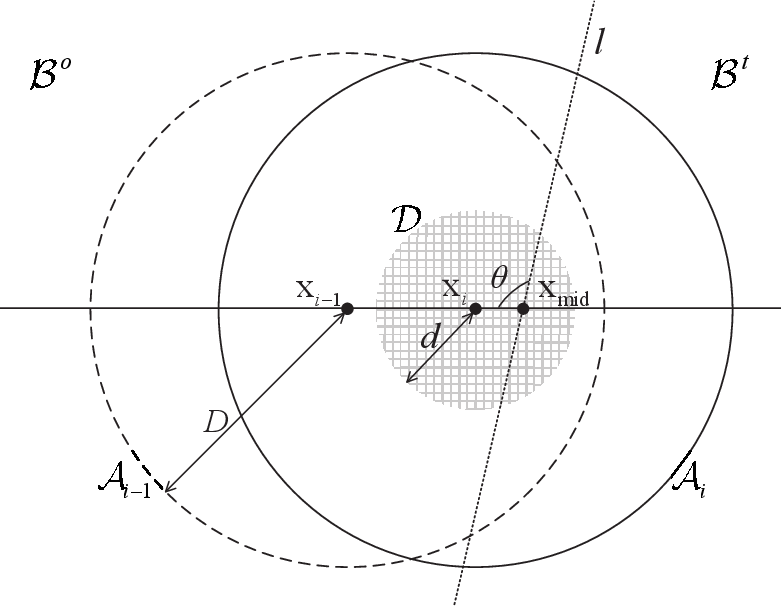}}
\vspace{-0.3cm}
\caption{Regions of interest in \emph{Proof of Lemma 1}, \emph{Lemma 3}, and \emph{Lemma 4}, where the dashed circle is the region $\mathcal{A}_{i-1}$, the solid circle is the region $\mathcal{A}_{i}$, the shaded circle is the region $\mathcal{D}$.}
\vspace{-0.6cm}
\label{fig1}
\end{figure}

\vspace{-0.4cm}
\subsection{Proof of Lemma 3}
In this proof, the regions and corresponding symbols defined in the proof of \emph{lemma 1} are used. Take the example of $k\!=\!o$ again. As shown in Fig. 13, the region centered on the user's location at the $i$-th measurement, where $d$ is the radius, is denoted as $\mathcal{D}$. The region where the IRS of original BS may exist with a distance from the user less than $d$ at the $i$-th measurement is denoted as ${\mathcal{D}}' = {\cal D} \cap {{\cal B}^o}$. As for the state ${{\mathcal I}_{2}^o}$. Then the cumulative distribution function (cdf) of $d$ under ${{\mathcal I}_{2}^o}$ is given by
\begin{equation}
\setlength\abovedisplayskip{3pt}
\setlength\belowdisplayskip{3pt}
\begin{small}
\begin{aligned}
{F_d}\left( {d\left| {i,{\cal I}_2^o} \right.} \right) =\! \frac{{{\mathbb P}\left( {N\left| {{\cal D}' \backslash \left( {{{\cal D}^\prime } \cap {\cal A}'_{i - 1} } \right)} \right| \!\ne\! 0} \right)}}{{{\mathbb P}\left( {N\left| {{{\cal A}^\prime }_i\backslash {\cal A}_{i,i - 1}^ \cap } \right| \!\ne\! 0} \right)}},
\end{aligned}
\end{small}
\end{equation}
where $\left| {{{\mathcal A'}_i}\backslash {\mathcal A}_{i,i - 1}^ \cap } \right| = \left| {{\mathcal A'}_i} \right| - \left| {\mathcal A}_{i,i - 1}^ \cap \right|$, $\left| {{\mathcal A'}_i} \right|$ and $\left| {\mathcal A}_{i,i - 1}^ \cap \right|$ are given in (42),  $\left| {{\cal D}' \backslash \left( {{{\cal D}^\prime } \cap {\cal A}'_{i - 1} } \right)} \right| = \left| {{\cal D}'} \right| - \left|  {{{\cal D}^\prime } \cap {\cal A}'_{i - 1} }   \right|$, $\left| {{\cal D}'} \right|$ and $\left|  {{{\cal D}^\prime } \cap {\cal A}'_{i - 1} }  \right|$ are given by
\begin{equation}
\begin{small}
\begin{aligned}
&\left| {{\cal D}'} \right| = \left\{ \begin{array}{l}
\left| {\cal D} \right|,\ \ \ \ \ \ \ \ \ \ \ \ \ 0 < d \le D   \wedge  0 \le {x_i} \le {x_{{\rm{mid}}}} - \frac{d}{{\sin \theta }}\\
\left| {\cal D} \right| - \left| {{\cal D} \cap {{\cal B}^t}} \right|,\\
\ \ \left| {{x_{{\rm{mid}}}} \!-\! {x_i}} \right|\sin \theta  \!<\! d \!\le\! D  \wedge {x_{{\rm{mid}}}} \!-\! \frac{D}{{\sin \theta }} \!<\! {x_i} \!\le\! {x_{{\rm{mid}}}} \!+\! \frac{D}{{\sin \theta }}\\
0,\ \ \ \ \ \ \ \ \ \ \ \ \ \ \  {x_{{\rm{mid}}}} + \frac{D}{{\sin \theta }} < {x_i}
\end{array} \right.,\\
&\left| {{\cal D}' \cap {{{\cal A}'}_{i - 1}}} \right| = \left\{ \begin{array}{l}
\left| {{\cal D} \cap {{\cal A}_{i - 1}}} \right|,\ \ \ 0 \le {x_i} \le {x_{{\rm{mid}}}} - \frac{D}{{\sin \theta }} + \Delta x\\
\left| {\left( {{\cal D} \cap {{\cal A}_{i - 1}}} \right)\backslash {{\cal B}^t}} \right|,\\
\ \ \ \ \ \ \ \ \ \ \ \ {x_{{\rm{mid}}}} \!-\! \frac{D}{{\sin \theta }} \!+\! \Delta x \!<\! {x_i} \!\le\! {x_{{\rm{mid}}}} \!+\! \frac{D}{{\sin \theta }}\\
0,\ \ \ \ \ \ \ \ \ {x_{{\rm{mid}}}} + \frac{D}{{\sin \theta }} < {x_i}
\end{array} \right.,
\end{aligned}
\end{small}
\end{equation}
By calculating the area of regions and substituting (42) and (44) into (43), the cdf is derived. Thus, the corresponding pdf is ${f_d}\left( {d\left| {i,{\cal{I}}_2^o} \right.} \right) = {{{\rm{d}}{F_d}\left( {d\left| {i,{\cal{I}}_2^o} \right.} \right)} \mathord{\left/
 {\vphantom {{{\rm{d}}{F_d}\left( {d\left| {i,I_2^o} \right.} \right)} {{\rm{d}}d}}} \right.
 \kern-\nulldelimiterspace} {{\rm{d}}d}}$, we obtain the pdf as (16).

For state ${{\mathcal I}_{4}^o}$, the cdf of $d$ under ${{\mathcal I}_{4}^o}$ is given by
\begin{equation}
\setlength\abovedisplayskip{3pt}
\setlength\belowdisplayskip{3pt}
\begin{small}
\begin{aligned}
F_{{i}}^{{\cal I}_{11}^o}\left( d \right) = \frac{{{\mathbb P}\left( {N\left| {\cal D}' \right| \ne 0} \right)}}{{{\mathbb P}\left( {N\left| {{{\cal A}^\prime }_i} \right| \ne 0} \right)}},
\end{aligned}
\end{small}
\end{equation}
By calculating the area of regions and substituting (42) and (44) into (43), the cdf is derived. Thus the corresponding pdf is ${f_d}\left( {d\left| {i,{\cal{I}}_4^o} \right.} \right) = {{{\rm{d}}{F_d}\left( {d\left| {i,{\cal{I}}_4^o} \right.} \right)} \mathord{\left/
 {\vphantom {{{\rm{d}}{F_d}\left( {d\left| {i,I_4^o} \right.} \right)} {{\rm{d}}d}}} \right.
 \kern-\nulldelimiterspace} {{\rm{d}}d}}$, we get the pdf in (16), where $\left| {{\cal D}'} \right|$ and $\left|  {{{\cal D}^\prime } \cap {\cal A}'_{i - 1} }  \right|$ are denoted as $S^o_{i,d}$ and $S^{o,\cap}_{i,d}$ for the sake of simplicity.

Using the same procedures, 
the pdfs of the distance between the user and serving IRS under ${\mathcal I}_{2}^t$ and ${\mathcal I}_{4}^t$ can be obtained, where $\mathcal{A}'_{i}$, $\mathcal{A}'_{i-1}$, and $\mathcal{D}'$ need to be modified to $\mathcal{A}_{i} \cap \mathcal{B}^t$, $\mathcal{A}_{i-1} \cap \mathcal{B}^t$, and $\mathcal{D} \cap {\mathcal{B}}^t$.

\vspace{-0.4cm}
\subsection{Proof of Lemma 4}
In this proof, the regions and corresponding symbols defined in the proof of \emph{lemma 1} and \emph{lemma 3}  are used. Take the example of $k\!=\!o$ again. The circle with the user's location at the $i$-th measurement as the center and $d$ as the radius, i.e., the boundary of $\mathcal{D}$, is denoted as $\mathcal{C}$. Therefore, for a given $d$, IRS may exist on ${\cal C}' = {\cal C} \!\cap\! {{\cal B}^o}$.

As for the state ${{\mathcal I}_{2}^o}$, the IRS exists on ${\cal C}' \backslash {{\cal A}_{i - 1}}$ with equal probability, and angles corresponding to the intersections of ${\cal C}'$ and ${{\cal A}_{i - 1}}$ are $\frac{\kappa }{2}$ and $2\pi - \frac{\kappa }{2}$, where $\kappa \! =\! 2\arccos \frac{{{D^2} - \Delta {x^2} - {d^2}}}{{2\Delta xd}}$. As for the state ${{\mathcal I}_{4}^o}$, the IRS exists on ${\cal C}'$ with equal probability, and angles corresponding to the intersections of ${\cal C}$ and ${{\cal B}^o}$ are $\frac{{\mu  + \pi }}{2} \!-\! \theta $ and $\frac{{5\pi  - \mu }}{2} - \theta $, where $\mu  = 2\arccos \frac{{\left| {{x_{{\rm{mid}}}} - {x_i}} \right|\sin \theta }}{d}$. Thus, the IRS is uniformly distributed over the angle range in which it may exist and the pdfs are given in (20). The pdfs under ${\mathcal I}_{2}^t$ and ${\mathcal I}_{4}^t$ can be obtained by modifying $\mathcal{C}'$ to $\mathcal{C} \cap {\mathcal{B}}^t$.

\end{appendix}

\vspace{12pt}

\end{document}